\documentclass[sigconf, authorversion]{acmart}
\AtBeginDocument{%
  }

\copyrightyear{2025}
\acmYear{2025}
\setcopyright{acmlicensed}
\acmConference[MM '25]{Proceedings of the 33rd ACM International Conference on Multimedia}{October 27--31, 2025}{Dublin, Ireland}
\acmBooktitle{Proceedings of the 33rd ACM International Conference on Multimedia (MM '25), October 27--31, 2025, Dublin, Ireland}
\acmDOI{10.1145/3746027.3755871}
\acmISBN{979-8-4007-2035-2/2025/10}

\usepackage{multirow, multicol}
\usepackage{booktabs, tabularx}
\usepackage{subcaption}
\usepackage{xcolor}

\usepackage{listings}
\usepackage{ifpdf}

\begin{document}

\title{SyMuPe: Affective and Controllable Symbolic Music Performance}


\author{Ilya Borovik}
\affiliation{%
  \institution{Skolkovo Institute of Science and Technology}
  \city{Moscow}
  % \state{Moscow}
  \country{Russia}
}
\email{ilya.borovik@skoltech.ru}
\orcid{0009-0002-1186-2039}

\author{Dmitrii Gavrilev}
\affiliation{%
  \institution{Skolkovo Institute of Science and Technology}
  \city{Moscow}
  % \state{Moscow}
  \country{Russia}
}
\email{dmitrii.gavrilev@skoltech.ru}
\orcid{0009-0005-6904-2852}

\author{Vladimir Viro}
\affiliation{%
  \institution{Peachnote GmbH}
  \city{Munich}
  % \state{Bavaria}
  \country{Germany}
}
\email{vladimir@peachnote.de}
\orcid{0000-0002-0473-0890}

% \renewcommand{\shortauthors}{Borovik et al.}
\renewcommand{\shortauthors}{Ilya Borovik, Dmitrii Gavrilev, and Vladimir Viro}

\begin{abstract}
Emotions are fundamental to the creation and perception of music performances. However, achieving human-like expression and emotion through machine learning models for performance rendering remains a challenging task. In this work, we present SyMuPe, a novel framework for developing and training affective and controllable symbolic piano performance models. Our flagship model, PianoFlow, uses conditional flow matching trained to solve diverse multi-mask performance inpainting tasks. By design, it supports both unconditional generation and infilling of music performance features. For training, we use a curated, cleaned dataset of 2,968 hours of aligned musical scores and expressive MIDI performances. For text and emotion control, we integrate a piano performance emotion classifier and tune PianoFlow with the emotion-weighted Flan-T5 text embeddings provided as conditional inputs. Objective and subjective evaluations against transformer-based baselines and existing models show that PianoFlow not only outperforms other approaches, but also achieves performance quality comparable to that of human-recorded and transcribed MIDI samples. For emotion control, we present and analyze samples generated under different text conditioning scenarios. The developed model can be integrated into interactive applications, contributing to the creation of more accessible and engaging music performance systems.
\end{abstract}

\begin{CCSXML}
<ccs2012>
   <concept>
       <concept_id>10010405.10010469.10010475</concept_id>
       <concept_desc>Applied computing~Sound and music computing</concept_desc>
       <concept_significance>500</concept_significance>
       </concept>
   <concept>
       <concept_id>10010520.10010521.10010542.10010294</concept_id>
       <concept_desc>Computer systems organization~Neural networks</concept_desc>
       <concept_significance>500</concept_significance>
       </concept>
   <concept>
       <concept_id>10002951.10003227.10003251</concept_id>
       <concept_desc>Information systems~Multimedia information systems</concept_desc>
       <concept_significance>300</concept_significance>
       </concept>
 </ccs2012>

\end{CCSXML}





\maketitle
\vspace{6pt}

\section{Introduction}\label{sec:introduction}

Music performance, a nuanced and unique interpretation and execution of a written music piece \cite{palmer1997music}, is deeply connected to emotions from the perspective of both the performer and the listener \cite{gabrielsson1996emotional, juslin2003communication}. These emotions are conveyed through subtle variations in note timing, dynamics, and articulation. One way to express emotion through music is to learn and perform it yourself. An alternative is to use a computational model \cite{cancino2018computational} that can simulate expressive performance and provide intuitive control over emotional intent. Despite advances in deep learning, existing models \cite{jeong2019graph, borovik2023scoreperformer, zhang2024dexter, zhang2025renderbox, tang2025towards} are either limited by data size, lack of human-level realism, intuitive control, or open implementations.

We address the controllable rendering of expressive music performances and present \textbf{SyMuPe}\footnote{Demo: \url{https://ilya16.github.io/SyMuPe}}, a flexible framework for developing and training affective and controllable \underline{Sy}mbolic \underline{Mu}sic \underline{Pe}rformance models. SyMuPe includes tools for preparing score-performance data, encoding it with a configurable SyMuPe tokenizer, and building varying transformer-based \cite{vaswani2017attention} encoder and decoder models. It supports a variety of learning objectives, including masked and causal language modeling and flow matching.

Advancing the research on symbolic piano performance modeling, we also present \textbf{PianoFlow}, a multimodal model that supports both unconditional and fine-grained performance rendering via text and emotion inputs. Built on conditional flow matching \cite{lipman2022flow}, PianoFlow enables high-quality performance generation with fast inference. Control is achieved by combining emotion classification and text-conditioned training using emotion-weighted embeddings from a pre-trained Flan-T5 encoder~\cite{chung2024scaling}.

PianoFlow is trained on a curated dataset of 2,968 hours of aligned score and performance MIDI pairs, surpassing publicly available alternatives. For emotion conditioning, we train a classifier on professionally recorded excerpts labeled by 33 emotions and use it to softly label the main dataset. PianoFlow is then fine-tuned with text embedding inputs, produced by weighting the embeddings for emotion description templates according to the emotion probabilities. After fine-tuning, the model can be controlled with any text prompt, including emotion labels or short descriptive phrases.

We evaluate PianoFlow against transformer-based baselines and external models using objective metrics and a blind, side-by-side listening test. The model achieves the best overall results, winning 67\% of all pairwise comparisons and 54\% when compared directly to human performances recorded as MIDI or transcribed from audio. The model responds to arbitrary textual input in an expected and musically coherent way.
With support for real-time inference, PianoFlow can be used in interactive applications for exploring piano music and creating custom music performances \cite{borovik2023nime, borovik2023hhai}. % ~\cite{borovik2023hhai}.

With this work, we contribute to the development of more intelligent and controllable models and tools for the performance of written musical compositions. Our main contributions are:

\begin{enumerate}
    \item \textbf{SyMuPe}, a flexible framework for tokenizing score and performance MIDI and creating transformer-based models for expressive music performance rendering.

    \item \textbf{PianoFlow}, a state-of-the-art model based on flow matching for controllable creation of expressive piano performances.

    \item \textbf{Multimodal performance rendering control} lying on the intersection of music and emotions and embedded using fine-grained emotion-weighted Flan-T5 text embeddings.
\end{enumerate}

\section{Related Work}

\subsection{Expressive Music Performance Rendering}

The field of expressive music performance rendering is dominated by machine learning methods. The temporal dependencies in a music performance are encoding using recurrent neural networks \cite{cancino2018thesis, jeong2019virtuosonet, maezawa2019rendering, rhyu2022sketching, renault2023expressive}, graph neural networks \cite{jeong2019graph}, transformers \cite{borovik2021thesis, borovik2023scoreperformer, tang2023reconstructing, worrall2024comparative, tang2025towards}, a U-Net \cite{zhang2024dexter}, and diffusion transformers (DiT) \cite{zhang2025renderbox}. Generative modeling paradigms include conditional variation autoencoders and autoregressive decoding \cite{maezawa2019rendering, jeong2019virtuosonet, jeong2019graph, borovik2021thesis, rhyu2022sketching, borovik2023scoreperformer}, generative adversarial networks \cite{renault2023expressive}, one-step performance prediction \cite{tang2023reconstructing, tang2025towards}, and diffusion \cite{zhang2024dexter, zhang2025renderbox}. Recent trends show a shift towards audio performance rendering \cite{tang2025towards, zhang2025renderbox, tang2025midivalle}, however, the models do not yet support interactive real-time inference.
For performance rendering control, the systems usually provide global \cite{jeong2019virtuosonet, tang2025midivalle} and fine-grained \cite{maezawa2019rendering, rhyu2022sketching} control of performance style, use perceptual performance features \cite{zhang2024dexter}, musical performance directions \cite{jeong2019virtuosonet, borovik2023hhai, borovik2023scoreperformer}, performer labels \cite{tang2023reconstructing, tang2025towards}, and textual descriptions \cite{zhang2025renderbox}.

\subsection{Music Performance and Emotions}

Music performance is deeply connected with emotions, both from the perspective of the performer and the audience \cite{gabrielsson1996emotional, lundqvist2009emotional}. For emotion-driven music generation \cite{hung2021emopia, bao2022generating, huang2024emotion}, emotions are usually represented using the two-dimensional framework of valence and arousal derived from Russell's Circumplex model of affect \cite{russell1980circumplex}. When discrete labels are used, the number of classes is typically limited to a small set (e.g., happy, sad, angry, calm) rather than dozens \cite{dash2024affective}. For piano expression, there are only proprietary datasets labeled with the discrete emotions \cite{hung2024eme33}. In our work, we cover a wide range of 33 emotions in piano performances in a curated recorded dataset, and generalize emotion control to arbitrary text using the embeddings of the pre-trained Flan-T5 text encoder \cite{chung2024scaling}.

\subsection{Flow Matching for Generative Modeling}

Flow matching \cite{lipman2024flow} has achieved state-of-the-art results in image and video generation \cite{batifol2025flux, stoica2025contrastive, jin2025pyramidal}, text-to-speech synthesis \cite{le2024voicebox, mehta2024matcha, chen2024f5, guo2024voiceflow}, and text-to-music generation \cite{prajwal2024musicflow, tal2024jasco, fei2024flux}. Unlike diffusion models \cite{yang2023diffusion}, which use iterative reverse diffusion processes, flow matching directly aligns the model's vector field with the target dynamics, leading to shorter sampling paths and faster inference. The application of flow matching to the rendering of expressive music performances is unexplored. In our work, we use it to capture the subtle nuances of piano performance, enabling high-fidelity and parallel, near-real-time controllable performance rendering.

\section{Background}
\label{sec:background}

\subsection{Music Performance Rendering}

Music performance rendering involves transforming symbolic score representations into expressive performances by simulating the nuances of a human performance. 
This process can be formalized as a conditional sequence generation task, where performance characteristics are predicted given musical score information and optional global or fine-grained control. Formally, data modalities are:

\begin{enumerate}
    \item a \textit{sequence of score note features} (note pitch, position, duration): $y = \{y_1, ..., y_n\}$, $y \in \mathbb{R}^{n \times s}$,
    \item a \textit{sequence of performance note features} (tempo and local timing, dynamics, articulation): $x = \{x_1, ..., x_n\}$, $x \in \mathbb{R}^{n \times p}$,
    \item optional \textit{conditioning embeddings} (performer style, direction markings, emotions or textual descriptions): $c \in \mathbb{R}^{m \times d}$
\end{enumerate}

The rendering task requires learning a mapping:
$x_t = f_\theta(x_{<t}, y, c)$ that predicts performance features \( x_t \) for each note using past performed notes \( x_{<t} \), score features \( y \) and condition \( c \). The models are trained to maximize the likelihood of observed performances:
\begin{equation}
\hat{\theta} = \underset{\theta}{\mathrm{argmax}} 
 \, p_\theta(x|y,c) = \underset{\theta}{\mathrm{argmax}} \prod_{t=1}^n p_\theta(x_t|x_{<t}, y, c)
\end{equation}

The factorization of the performance note feature distribution can be learned using an autoregressive (note-by-note) \cite{jeong2019virtuosonet, jeong2019graph, rhyu2022sketching, borovik2023scoreperformer} and parallel (all-at-once) \cite{tang2023reconstructing, zhang2024dexter, tang2025towards} generative modeling paradigms.

\subsection{Score and Performance Data}
\label{subsec:score-perf-data}

For symbolic music performance, the score is available in MusicXML \cite{good2001musicxml} and MIDI formats. The performances are either recorded as MIDI or transcribed from audio. A critical component is the correct matching of scores and performances at the global and local levels. For symbolic expressive music performance, the models use note-to-note aligned score and performance sequences. The alignment is computed using algorithms based on Hidden Markov Models \cite{nakamura2017alignment} and Dynamic Time Warping \cite{peter2023parangonar, peter2023nasap}. The note aligner classifies the notes as matched, missing or inserted in performance.
% The alignment produces links of three types: a match, a missing score note, or an inserted performance note.

The alignments are not perfect and may contain errors, such as close score/performance notes being aligned with distant and unrelated performance/score notes. For music performance rendering the common approaches are to take only the aligned score and performance notes \cite{rhyu2022sketching, zhang2024dexter, tang2025towards}, clean up the deviating onsets \cite{xia2016expressive, jeong2019virtuosonet, borovik2023scoreperformer}, and optionally interpolate the unperformed notes \cite{borovik2023scoreperformer}. 

\ifpdf
\begin{figure*}[!ht]
	\centering
    \includegraphics[width=1.\textwidth]{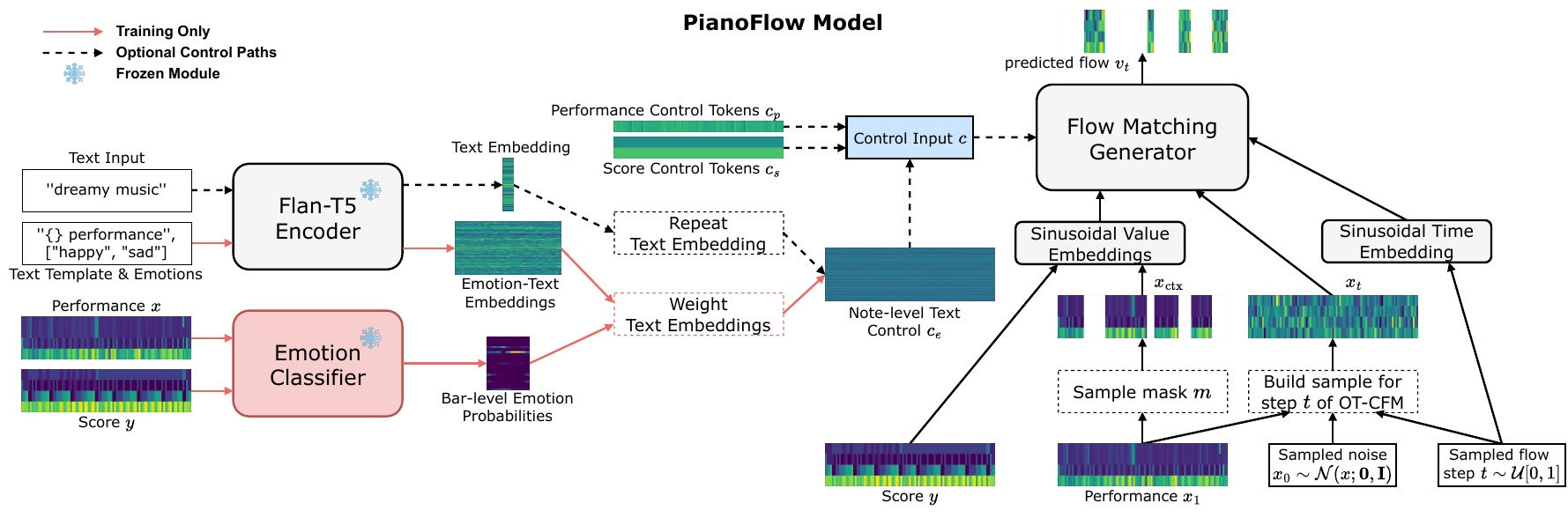}
	\caption{Architecture of the PianoFlow model with the control inputs on the left and generation logic on the right.}
	\label{fig:model}
\end{figure*}
\else\fi

\subsection{Flow Matching and OT-CFM}
\label{subsec:flow_matching}

Flow Matching (FM)~\cite{lipman2022flow, lipman2024flow} is a continuous framework for transforming one probability distribution into another through a learned time-dependent vector field.
Consider a target distribution $q(x)$ on $\mathbb{R}^d$ and a simple prior $p_0(x)$ (usually, $p_{0} \sim \mathcal{N}(x;\mathbf{0},\mathbf{I})$). The goal is to evolve $p_0$ into $p_1 \approx q$ via a flow $\phi_t(x): [0,1] \times \mathbb{R}^d \to \mathbb{R}^d$, which is defined by the ordinary differential equation (ODE):
\begin{equation}
    \frac{d}{dt}\phi_t(x) = u_t\bigl(\phi_t(x)\bigr), 
    \quad \phi_0(x) = x, 
    \quad t \in [0,1],
    \label{eq:flow_ode}
\end{equation}
where $u_t(x): [0,1] \times \mathbb{R}^d \rightarrow \mathbb{R}^d$ is a vector field that defines the dynamics of individual points from $t=0$ to $t=1$.

The fundamental learning objective is to match the parameterized vector field $v_t(x;\theta)$ to a target vector field $u_t(x)$ that generates the desired probability path. This is formalized by the FM objective:
\begin{equation}
    \mathcal{L}_{\text{FM}}(\theta) 
    = \mathbb{E}_{t \sim \mathcal{U}[0,1],\, x \sim p_t(x)}
    \bigl\|u_t(x) - v_t(x; \theta)\bigr\|^2
    \label{eq:fm_loss}
\end{equation}

In practice, direct computation of this objective is intractable because it requires knowledge of the marginal $p_t(x)$ and vector field $u_t(x)$. Conditional Flow Matching (CFM) \cite{lipman2022flow} solves this by decomposing the problem into conditional paths $p_t(x|x_1)$ for individual data points $x_1 \sim q$, which leads to the tractable objective:
\begin{equation}
    \mathcal{L}_{\text{CFM}}(\theta) = \mathbb{E}_{t \sim \mathcal{U}[0,1],x_1 \sim q, x \sim p_t(x|x_1)} \|u_t(x|x_1) - v_t(x;\theta)\|^2
    \label{eq:cfm_loss}
\end{equation}

FM and CFM have identical gradients with respect to $\theta$~\cite{lipman2022flow}, making CFM a valid and practical alternative for learning flows. 

A crucial design choice in CFM is the definition of the conditional paths $p_t(x \mid x_1)$. An effective option is the optimal transport (OT) path, a simple linear interpolation between a sample $x_0$ from $p_0$ and a data point $x_1$ from $q$:
\begin{align}
    \phi_t^{\text{OT}}(x) &= (1 - (1 - \sigma_{\min})t)x_0 + t x_1,
    \label{eq:ot_path} \\
    u_t^{\text{OT}}(x|x_1) &= x_1 - (1 - \sigma_{\min})x_0
    \label{eq:ot_field}
\end{align}

\section{Methodology}
\label{sec:methods}

\subsection{Data Encoding}
\label{sec:data-encoding}

For representing the score and performance sequences we propose a unified data encoding which consists of four score and four performance features. The score features $y_i \in \mathbb{R}^{4}$ include:
\begin{enumerate}
    \item \textbf{Pitch}: an integer MIDI pitch value in the range $[21, 108]$;
    \item \textbf{Position}: a position of the note inside the bar, with support for bars up to 2 whole note length and minimum resolution of 1/96th note to correctly encode triplets and short notes;
    \item \textbf{PositionShift}: an inter-note-interval between consecutive score notes, encoded with the same precision as positions. For chords, all notes have a zero shift except the first note;
    \item \textbf{Duration}: a score note value, scaled by the whole note. 
\end{enumerate}

The performance features $x_i \in \mathbb{R}^{4}$ include:
\begin{enumerate}
    \item \textbf{Velocity}: an integer MIDI velocity value in the range $[0, 127]$;
    \item \textbf{TimeShift}: an inter-note-interval between neighboring notes, encoded in seconds. It represents the local tempo between score onsets and inter-onset timing and deviations.
    \item \textbf{TimeDuration}: a performed (pressed) note duration, encoded in seconds.
    \item \textbf{TimeDurationSustain}: a performed note duration with the account for the pressed sustain pedal. It is always equal or larger than the explicit performed time duration.
\end{enumerate}

Unless explicitly stated, we use real-valued features, following the literature \cite{cancino2018thesis, jeong2019score, jeong2019virtuosonet, zhang2024dexter}. This avoids the loss of expressive detail caused by quantization in MIDI-based encodings \cite{huang2020pop, zeng2021musicbert, hsiao2021compound} and some rendering models \cite{borovik2023scoreperformer, tang2025towards}. MIDI pitches and velocities are normalized to $[0, 1]$, score positions and durations are scaled by the whole note, performance timing is kept in seconds.

Data sequences are sorted first by score time, and then by pitch for notes in the same chord. As a result, the time shift values may be negative because the micro-timing performed for the close notes may not follow the sorting of the notes in the score.

Pedal effects are crucial in piano performance. Prior work \cite{jeong2019virtuosonet, jeong2019graph, zhang2024dexter} has approximated the times and values of sustain pedals at the note boundaries. In reality, pedals form a separate sequence of timed features that are not aligned with exact notes. Here, we approximate the effect of sustain pedals by using a factorized representation for both pressed and sustained note durations. In the context of note-aligned scores and performances, this is a reasonable simplification.

\subsection{PianoFlow}
\label{subsec:model}

In this section we present PianoFlow, a transformer-based \cite{vaswani2017attention} model for expressive piano performance rendering utilizing the conditional flow matching (CFM) generative paradigm \cite{lipman2022flow}. The complete model with all conditioning is illustrated in Figure~\ref{fig:model}.

\ifpdf\else
\begin{figure*}[!ht]
	\centering
    \includegraphics[width=1.\textwidth]{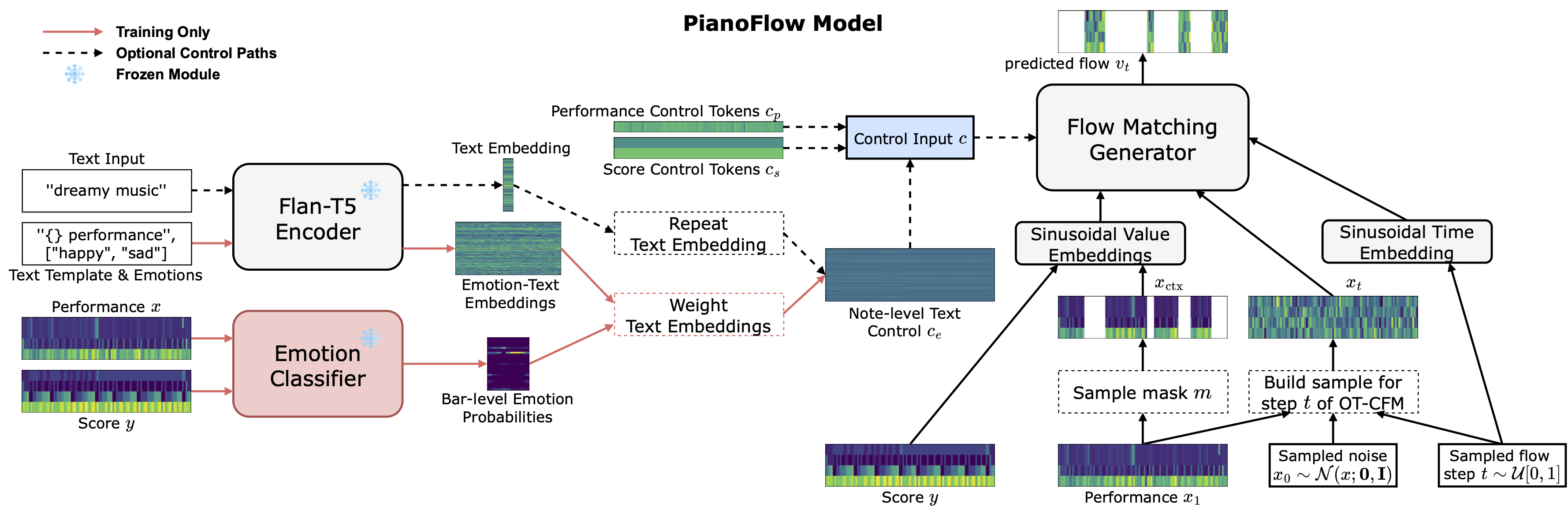}
	\caption{Architecture of the PianoFlow model with the control inputs on the left and generation logic on the right.}
	\label{fig:model}
\end{figure*}
\fi

% \subsubsection{Problem Formulation}
We formulate the generative task as a performance feature inpainting (unmasking) task. Let $m \in \{0, 1\}^{n}$ be a binary temporal mask of the same length as the note sequences $x$ and $y$. Then, $x_{\mathrm{m}} = m \odot x$ and $x_{\mathrm{ctx}} = (1 - m) \odot x$ are masked and known performance feature sequences, respectively. The generator learns $p_\theta (x_{\mathrm{m}} | x_{\mathrm{ctx}}, y)$, a distribution of performance note features given score and known performance feature context.

In the context of CFM \cite{lipman2022flow}, the model learns a time-dependent conditional vector field $v_t: [0, 1] \times \mathbb{R}^p \times \mathbb{R}^{p+s} \rightarrow \mathbb{R}^p$ used to construct a probability flow $\phi_t: [0, 1] \times \mathbb{R}^p \times \mathbb{R}^{p+s} \rightarrow \mathbb{R}^p$ from the prior distribution $p_0 \sim \mathcal{N}(x;\mathbf{0},\mathbf{I})$ to the target distribution $p_1 \sim q(x)$. The model is optimized only on the prediction of masked features. The masked CFM objective is:
\begin{equation}\label{eq:model-loss}
\mathcal{L}_{\text{CFM-m}}(\theta) = \mathbb{E}_{t,m,q(x),p_0} || m \odot (u(x_t|x) - v_t(x_t, x_{\mathrm{ctx}}, y; \theta)) ||^2
\end{equation}

% \subsubsection{Model Architecture}

The PianoFlow model consists of a transformer encoder \cite{vaswani2017attention} with the hidden dimension $D$. In attention layers, we use single key-value attention heads \cite{shazeer2019fast} for faster decoding and rotary positional embeddings \cite{su2024roformer} for encoding relative position information. In feedforward layers, we use SwiGLU activation \cite{shazeer2020glu}. To embed the flow time step, we use a sinusoidal position encoding mapping $t \in [0, 1]$ to $h_t \in \mathbb{R}^{D_t}$. These embeddings are added to the adaptive layer normalization layers before attention and feedforward layers. 

As described in Section~\ref{sec:data-encoding}, the score and performance features $y$ and $x$ are represented as vectors. In addition, we use special tokens for sequence boundaries (SOS and EOS) and MASK tokens for the inpainting. For the context of scores $y$ and performances $x_{\mathrm{ctx}}$, we combine discrete and sinusoidal embeddings \cite{guo2023fme, borovik2023scoreperformer}. Special tokens are encoded using a learned embedding table, while note features are embedded using a sinusoidal embedding followed by a linear projection. This method preserves the continuity and relative information between musical attributes (pitch, velocity, time and duration) that is lost in quantized discrete token representations.

\subsection{Training and Inference}
\label{subsec:training}

The score-only PianoFlow model is trained on a dataset of note-aligned scores and performances by optimizing the masked CFM objective in~\eqref{eq:model-loss}. Random flow step $t \sim \mathcal{U}[0, 1]$ and noise $x_0 \sim \mathcal{N}(x;\mathbf{0},\mathbf{I})$ are sampled to produce a noisy performance feature vector $x_t = (1 - (1 - \sigma_{\min})t)x_0 + t x_1, x_t \in \mathbb{R}^{n \times p}$. The sampled mask $m$ is used to construct the context $x_{\mathrm{ctx}}$. These inputs and score features $y$ are used to output and learn the vector field $v_t(x_t, x_{\mathrm{ctx}}, y; \theta)$.

During training, we sample different masks $m$ for performance features $x$ to make the model as general as possible. We call this multi-mask training. We design 5 different masks covering random notes, beats, bars, continuous segments, and the end of the sequence. The first four note masks train the model to inpaint random parts of the performance. The end-of-sequence mask trains the model to continue the performance using the past context. The masks are sampled uniformly and constitute half of the batch sequences. For the other half, we mask all performance features to train the model for a fully unconditional performance rendering. 

During inference, we encode musical scores and feed them to the model for unconditional rendering of music performance. The entire performance is predicted using overlapping windows in which K last notes are played by the model.
To sample from the learned audio distribution $p_1(x \mid y, x_{\text{ctx}})$, a noise $x_0$ is sampled from $p_0 \sim \mathcal{N}(x;\mathbf{0},\mathbf{I})$ and an ODE solver is used to evaluate $\phi_1(x_0)$ given $\frac{d}{dt}\phi_t(x) = v_t(\phi_t(x), x_{\text{ctx}}, y; \theta)$ and the initial condition $\phi_0(x_0) = x_0$. The ODE is solved by a number of function evaluations (NFE) by iterative integration from $t = 0$ to $t = 1$. The number of integration steps $k$ offers a balance between quality and speed of decoding.

\subsection{Emotion Classifier}
\label{subsec:emotion-classifier}

% The previous section described score-only piano performance rendering model. 
For interactive inference, we use emotions as intuitive control labels. 
First, we tackle the problem of emotion-driven performance classification. We pretrain a separate transformer encoder with the masked language modeling objective \cite{zeng2021musicbert} on the same data that we train all performance rendering models. Unlike performance rendering, we randomly mask both score and performance tokens. The training is multi-task and multi-mask. The tasks include: score encoding (no performance tokens), aligned performance encoding (all score and performance tokens), and time only performance encoding (no score tokens). Similar to performance rendering, the masks cover random notes, beats, bars, and segments.

After training a MLM backbone, we train a classifier using the embeddings of the penultimate embedding layer. The classifier consists of a single transformer layer that aggregates note embeddings into a sequence embedding that is linearly mapped to emotion probabilities. The classifier is trained on a curated and diverse dataset of short piano excerpts performed with 33 different emotions. The data is presented in Section~\ref{subsec:dataset}. The trained classifier is used to classify and label all performed musical bars in the main training dataset. The softly labeled bar-level emotion probabilities are used as a control signal for the generative model.

\subsection{Expression, Emotion and Text Conditioning}

We extend the described PianoFlow model for conditional and interactive inference. The control inputs include tempo and velocity score tokens, local performance tempo tokens and localized text embeddings for emotion inputs.

For score tokens $c_s$, we take the tempos and velocities associated with performance direction markings in MusicXML files. They tell the performer what tempo and dynamics to follow during the performance. For the model, these conditions put the generation within the range of expected dynamics and tempos used in human performances. We use these tokens by default in our models.

The performance tempo tokens $c_p$ are computed as local beat tempos for each onset in the score. These tokens can be used during inference to adjust the performance tempo.

Finally, for the text-based control, we use the text embeddings $c_e$ associated with each performed musical bar. We design a set of 16 text templates that can describe the emotions in music, e.g., \texttt{`[emotion]`}, \texttt{`[emotion] music`}, \texttt{`perform [emotion] music`}. The templates are sampled uniformly during training for each sequence and regularize the emotion-text space. We compute the average sentence embeddings for all 33 emotion labels and 16 templates using the pre-trained Flan-T5-base text encoder \cite{chung2024scaling}. We then take the emotion probabilities predicted by the trained emotion classifier, and for each bar we compute a weighted text embedding using the probabilities and emotion-template embeddings. Each note is assigned with a bar-level emotion-weighted text embedding. 

All three types of conditional inputs are concatenated and projected to produce an overall control input $c$. These embeddings are added to the hidden states before the middle transformer layer in the model. In this way, the first transformer layers explicitly focus on unconditional generation and then consider control. The experiments showed that the model converges faster with this form of conditioning than with the concatenation of control and inputs.

We tune the PianoFlow model with the control inputs. The whole setup is trained with a classifier-free guidance \cite{ho2022cfg}. During training, each conditional input type is dropped with individual probabilities. The training objective for the controllable PianoFlow model is:
\begin{equation}\label{eq:model-loss-cond}
\mathcal{L}_{\text{CFM-m-c}}(\theta) = \mathbb{E}_{t,m,q,p_0} || m \odot (u(x_t|x) - v_t(x_t, x_{\mathrm{ctx}}, y, c; \theta)) ||^2
\end{equation}

During inference, the conditioned vector field $\tilde{v}_t$ for the model incorporates guidance through interpolation:
\begin{equation}
    \tilde{v}_t(x_t, x_{\text{m}}, y, c; \theta) = \alpha v_t(x_t, x_{\text{ctx}}, y, c; \theta) + (1 - \alpha)v_t(x_t, x_{\text{ctx}}, y; \theta)
\end{equation}
where $\alpha$ controls the guidance strength.
% ($\alpha = 0$ is unconditional inference).

\subsection{SyMuPe: Tokenizer and Models}
\label{subsec:symupe}

PianoFlow model is a part of SyMuPe, a Symbolic Music Performance modeling framework. It includes the designed SyMuPe tokenizer and multiple transformer-based models. The framework is instrument agnostic and can be used for multi-instrumental pieces.

The SyMuPe tokenizer, built using \texttt{MidiTok} library \cite{fradet2021miditok}, supports a combined encoding and decoding of score and performance MIDI. The performance MIDI can be encoded using a time-only and a score-aligned encodings. The tokenizer supports both real-valued features and discrete tokens for regression and categorical feature prediction tasks.
For the tokenized representation, the score and performance features are quantized. For pitch and velocity, the MIDI values are used. The position, shift and duration tokens are quantized using an adaptive scheme with higher precision for short durations and shifts and lower precision for larger values \cite{zeng2021musicbert, fradet2021miditok}. 
Also, the tokenizer supports other tokens such as bar index, time signature (beat duration and max bar position), beat/bar tempo, pitch octave and class, number of notes and note index in the onset.

For models, the SyMuPe package allows the construction of encoder-only, decoder-only, and encoder-decoder transformer-based models for unconditional and conditional expressive music performance generation. The supported generative training paradigms are causal language modeling (CLM), masked language modeling (MLM), and conditional flow matching (CFM). We consider two baseline models in the experimental setup in Section~\ref{subsec:exp-model}.

\section{Experimental Setup}

\subsection{Dataset}
\label{subsec:dataset}

For our experiments, we clean and combine open-source datasets with our own curated piano score and performance data. We use the (n)ASAP \cite{foscarin2020asap, peter2023nasap} and ATEPP \cite{zhang2022atepp} datasets. (n)ASAP provides 92 hours of professional expressive piano performance MIDI aligned with musical scores at the beat and note level, while ATEPP contains around 1000 hours of transcribed piano performances, with 400 hours originally matched to musical scores.
Additionally, we collect more solo piano audio from the web using IMSLP composition title search and transcribe it to MIDI using the \texttt{Transkun V2} model \cite{yan2024transkun}.

The open-source data includes issues such as corrupted time signatures, empty MIDI notes, and duplicate compositions. We address these by recomputing musical scores using the \texttt{partitura} \cite{cancino2022partitura} library, fixing metadata errors in ATEPP, matching compositions from ASAP and ATEPP, and adding more matched scores from the PDMX dataset \cite{long2025pdmx}. The proprietary performances are matched with scores from ASAP, ATEPP, and PDMX.

All score and performance MIDI files are note-aligned and cleaned to produce the parallel score and performance note sequences. For the alignment, we use \texttt{Parangonar} \cite{peter2023parangonar}. During the alignment cleanup, we clean notes with high inter-onset and intra-onset deviations, process alignment holes and interpolate unperformed notes using local performance data. After the refinement, we obtain a parallel dataset of aligned score and performance note features.

Table \ref{tab:training-datasets} summarizes the data statistics. The combined PERiScoPe (Piano Expression Refined Score and Performance) dataset contains 36,977 piano performances (2,968 hours) aligned with 1,176 pieces by 67 composers, covering various historical periods and styles. Only 7.5\% of the notes are interpolated. This comprehensive, multi-source dataset allows to learn diverse performance characteristics. We use 90\% of the data for training and 10\% for validation and testing. To avoid data leakage, each musical composition and all its movements and performances appear in exactly one data split.

\begin{table}[t]
    \caption{Utilized score and performance datasets. \textit{Comp}, \textit{Score} and \textit{Perf} -- number of composers, scores, and performances, \textit{P/S} -- number of performances per score, \textit{Size} -- total duration in hours, \textit{Notes} -- total number of performed notes.}
    \label{tab:training-datasets}
    \newcolumntype{C}{>{\centering\arraybackslash}X}%
    \newcolumntype{L}{>{\raggedright\arraybackslash}X}%
    \begin{tabularx}{\columnwidth}{lCCCCCCC}
        \toprule
        \textbf{Dataset} & \textbf{Comp} & \textbf{Score} & \textbf{Perf} & \textbf{P/S} & \textbf{Size} & \textbf{Notes} \\
        \midrule
        (n)ASAP \cite{peter2023nasap} & 16 & 215 & 1004 & 4.7 & 89h & 3.4M \\
        ATEPP \cite{zhang2022atepp} & 14 & 338 & 5363 & 15.9 & 527h & 17.0M \\
        Proprietary & 67 & 1162 & 28179 & 24.3 & 2215h & 68.7M \\
        \midrule
        \textbf{PERiScoPe} & \textbf{67} & \textbf{1176} & \textbf{36977} & \textbf{31.5} & \textbf{2968h} & \textbf{93.1M} \\
        \bottomrule
    \end{tabularx}
\end{table}

For emotion classification, in the absence of annotated score performance data, we hired professional classically trained musicians to record an internal dataset of short affective music excerpts. Each excerpt, either a few bars of classical music or a short improvised sequence, is labeled with one of 33 emotions (e.g., \emph{happy}, \emph{calm}, \emph{sleepy}, \emph{comical}, \emph{anger}, \emph{anxious}, \emph{rapidly}, \emph{decisive}). % available on the demo web page.
% in the supplementary material.

\subsection{Model and Training Configurations}
\label{subsec:exp-model}

For \textbf{PianoFlow} and our baseline models, we use the following base configuration. The transformer has 8 layers of dimension 512 and 8 attention heads. The feedforward layer dimensions are set to 512 and 1536. The embedding dimensions for all features and sinusoidal time embedding are set to 64. Score tempo and velocity conditioning is embedded using a projection in the center of the transformer (before the 5th layer). For text-conditioned models, the Flan-T5 embeddings are also added to the same layer. Classifier-free guidance dropout probabilities for control inputs are set to 0.2 for each input type. During inference, we solve ODE using $k=10$ sampling steps with adaptive step size: $\Delta t_{i, i+1} = \gamma \Delta t_{i-1, i}, \gamma = 0.75$.

For the baselines, we take two transformer-based models built using SyMuPe framework presented in Section~\ref{subsec:symupe}. First, an encoder transformer with MLM loss (\textbf{MLM}) similar to the MIDI-to-MIDI module of the MIDI-to-Audio piano performance rendering model \cite{tang2025towards}. Second, an encoder-decoder trained with the causal language modeling loss for the decoder (\textbf{EncDec}). This model is a simplified style-unconditional version of the ScorePerformer \cite{borovik2023scoreperformer}. The models have a similar number of parameters as PianoFlow.

The maximum sequence length during training is set to 256 notes, and the batch size is set to 128. We randomly sample augmentations for training sequences with a probability of 0.5. Augmentations include pitch shift (up to $\pm 6$ semitones), velocity shift (up to $\pm 6$ MIDI bins), and tempo change ($\pm 0.05\%$). The ratio of masked notes is sampled from $[0.1, 0.9]$. We use the Adam optimizer \cite{kingma2014adam} with a weight decay of $1 \cdot 10^{-2}$ and a learning rate of $2 \cdot 10^{-4}$, which decays by half by the end of training (300,000 iterations). 
% Models are trained on the V100 GPU in a single day.

For external models, we compare with the existing state-of-the-art models. We take the VirtuosoNet model \cite{jeong2019virtuosonet, jeong2019graph} with an iterative sequential graph network (ISGN) (\textbf{VirtuosoNet}) and the DExter diffusion-based model \cite{zhang2024dexter} (\textbf{DExter}) with the only available checkpoint with score only conditioning. We use the official checkpoints provided with the code and make no changes to the inference code.

\subsection{Objective Evaluation}

Expressive music performance rendering lacks established and robust objective evaluation metrics. In our work, we follow the literature and use Pearson's correlation \cite{jeong2019graph, rhyu2022sketching, borovik2023scoreperformer} and KL divergence \cite{zhang2024dexter, tang2025towards}. The correlation evaluates the temporal change of the performance expression, while the divergence, computed using Monte Carlo estimation, compares the sequence-wise performance feature distributions. We evaluate:
\begin{enumerate}
    \item \texttt{Vel}: onset velocity for the dynamics curve;
    \item \texttt{IOI}: relative inter-onset intervals as a proxy for tempo, computed as the ratio of time and score onset shifts;
    \item \texttt{OD}: relative intra-onset deviations for local timing deviation, time deviations scaled by score onset shifts;
    \item \texttt{Art} and \texttt{ArtS}: note articulation, a ratio between performance and score note durations without and with sustain.
\end{enumerate}

For the repertoire, we take 80 musical scores from the test set that have at least 2 real performances in the dataset. Then, for each score, we synthesize 11 random performances from scratch. We compute the metrics between all pairs of performances in two distinct sets (e.g., real and generated) and average the metrics for each score. Finally, an overall metric is computed by averaging all per-score metrics. Thus, each musical composition contributes equally to the evaluation, regardless of how many real performances it has. 
% in the dataset.

\subsection{Listening Test}

We conduct a subjective evaluation to compare samples generated by PianoFlow, baseline and external models, together with the real performances from the dataset and deadpan (inexpressive) renditions. For real performances, we use recoded MIDI files from ASAP when available, and transcribed MIDI files from ATEPP otherwise. Deadpan samples are included to trick the listener and to assess the preference for expressive performances over raw score renderings.

For external models such as VirtuosoNet and DExter, we use the official trained checkpoints and inference code. As their training splits are not available, it is possible that some samples from ASAP and ATEPP were seen by the models. While retraining these models on our dataset would be ideal, adapting to their specific preprocessing and training pipelines is non-trivial. As part of our contributions, we provide a curated dataset and a universal tokenizer, which facilitate the development of simpler, more efficient models compared to existing feature-engineered or slower alternatives.

We evaluate the models on 20 classical piano excerpts from the Baroque, Classical, Romantic, and Impressionist periods. Each model performs every piece three times, and selected bars are cut for evaluation. The dataset samples are chosen randomly. All MIDI segments are synthesized with \texttt{Pianoteq}\footnote{\url{https://www.modartt.com/pianoteq_overview}} to ensure consistent acoustic quality. Excerpts range from 15 to 45 seconds. Evaluation is done using a custom Telegram bot that rotates audio samples across models and pieces. Participants rate samples based on expressiveness and musical quality, casting at least 25 votes each.

\section{Results and Discussion}

\subsection{Objective Evaluation}

Table~\ref{tab:objective-evaluation} summarizes the objective evaluation results. The correlation values indicate how closely each model follows the temporal feature contours of real performances. The first row (\texttt{Dataset}) shows typical agreement between human performances and serves as a reference and theoretical target for the models. \texttt{PianoFlow} achieves higher correlations for \texttt{IOI}, \texttt{OD}, and \texttt{Art}, indicating better modeling of timing variations than token-based baselines and prior models. For dynamics (\texttt{Vel}), \texttt{VirtuosoNet} shows a higher average correlation with real performances, although the per-score deviations are also higher. Transformer-based baselines show higher correlation with the evaluation set performances than the external models, suggesting that a more diverse training dataset benefits the models in generalizing to unseen music, even if external models may have seen some test pieces during training.

\begin{table*}[!ht]
    \caption{Statistical comparison of the rendered and human performances. \texttt{Vel} - velocity, \texttt{IOI} - inter-onset-interval, \texttt{OD} - relative onset deviation, \texttt{Art} - articulation, \texttt{ArtS} - sustained articulation. 
    Bold and underlined mark the best and second best models. \texttt{Param} - number of parameters, \texttt{NPS} - average number of notes rendered per second on a V100 GPU.}
    \label{tab:objective-evaluation}
    \newcolumntype{C}{>{\centering\arraybackslash}X}%
    \newcolumntype{L}{>{\raggedright\arraybackslash}X}%
    \resizebox{\textwidth}{!}{\begin{tabularx}
        {\textwidth}{lCCcCCCCCcCCCCC}
        \toprule
        &&&& \multicolumn{5}{c}{\textbf{Pearson’s Correlation ($\uparrow$)}} && \multicolumn{5}{c}{\textbf{KL Divergence ($\downarrow$)}} \\
        \cmidrule{5-9} \cmidrule{11-15}
        \textbf{Model} & \textbf{Param} & \textbf{NPS} && \textbf{Vel} & \textbf{IOI} & \textbf{OD} & \textbf{Art} & \textbf{ArtS} && \textbf{Vel} & \textbf{IOI} & \textbf{OD} & \textbf{Art} & \textbf{ArtS} \\
        \midrule
        Dataset & - & - && 0.65{\scriptsize $\pm$0.18} & 0.90{\scriptsize $\pm$0.11} & 0.24{\scriptsize $\pm$0.18} & 0.50{\scriptsize $\pm$0.14} & 0.46{\scriptsize $\pm$0.20} & & 0.27{\scriptsize $\pm$0.34} & 0.20{\scriptsize $\pm$0.25} & 0.13{\scriptsize $\pm$0.12} & 0.12{\scriptsize $\pm$0.10} & 0.15{\scriptsize $\pm$0.13} \\
        \midrule
        VirtuosoNet \cite{jeong2019graph} & 5M & 1700 && \textbf{0.48{\scriptsize $\pm$0.27}} & 0.58{\scriptsize $\pm$0.29} & 0.00{\scriptsize $\pm$0.05} & 0.27{\scriptsize $\pm$0.17} & - & & 0.73{\scriptsize $\pm$0.76} & 0.93{\scriptsize $\pm$1.08} & 1.57{\scriptsize $\pm$1.68} & 1.53{\scriptsize $\pm$1.47} &  - \\
        DExter \cite{zhang2024dexter} & 62M & 30 && 0.24{\scriptsize $\pm$0.30} & 0.36{\scriptsize $\pm$0.35} & 0.01{\scriptsize $\pm$0.04} & 0.14{\scriptsize $\pm$0.19} & - & & 1.67{\scriptsize $\pm$1.11} & \underline{0.47{\scriptsize $\pm$0.38}} & 0.45{\scriptsize $\pm$0.33} & 0.81{\scriptsize $\pm$0.52} &  - \\
        \midrule
        MLM & 24M & 3450 && 0.41{\scriptsize $\pm$0.18} & \underline{0.80{\scriptsize $\pm$0.13}} & \underline{0.07{\scriptsize $\pm$0.07}} & 0.30{\scriptsize $\pm$0.10} & 0.30{\scriptsize $\pm$0.14} & & \textbf{0.38{\scriptsize $\pm$0.45}} & \textbf{0.31{\scriptsize $\pm$0.22}} & \underline{0.19{\scriptsize $\pm$0.10}} & \textbf{0.23{\scriptsize $\pm$0.30}} & \textbf{0.30{\scriptsize $\pm$0.37}} \\
        EncDec & 24M & 80 && 0.37{\scriptsize $\pm$0.15} & \underline{0.80{\scriptsize $\pm$0.13}} & \underline{0.07{\scriptsize $\pm$0.09}} & 0.31{\scriptsize $\pm$0.11} & 0.32{\scriptsize $\pm$0.14} & & \underline{0.57{\scriptsize $\pm$0.63}} & 0.67{\scriptsize $\pm$0.86} & \textbf{0.16{\scriptsize $\pm$0.11}} & 0.47{\scriptsize $\pm$0.61} & \underline{0.35{\scriptsize $\pm$0.44}} \\
        \midrule
        \textbf{PianoFlow} & 24M & 355 && \underline{0.43{\scriptsize $\pm$0.17}} & \textbf{0.86{\scriptsize $\pm$0.12}} & \textbf{0.09{\scriptsize $\pm$0.12}} & \textbf{0.35{\scriptsize $\pm$0.12}} & \textbf{0.35{\scriptsize $\pm$0.17}} & & 0.64{\scriptsize $\pm$0.56} & 0.56{\scriptsize $\pm$0.76} & 0.55{\scriptsize $\pm$0.51} & \underline{0.46{\scriptsize $\pm$0.50}} & 0.48{\scriptsize $\pm$0.89} \\
        \quad w/o $c_s$ & 24M & 355 && 0.35{\scriptsize $\pm$0.16} & 0.84{\scriptsize $\pm$0.13} & 0.09{\scriptsize $\pm$0.13} & \underline{0.32{\scriptsize $\pm$0.11}} & \underline{0.34{\scriptsize $\pm$0.17}} & & 0.74{\scriptsize $\pm$0.59} & 0.99{\scriptsize $\pm$1.14} & 0.56{\scriptsize $\pm$0.53} & 0.75{\scriptsize $\pm$0.70} & 0.75{\scriptsize $\pm$1.14} \\
        \bottomrule
    \end{tabularx}}
\end{table*}

The KL divergence scores (right panel) measure how closely the model's global feature distribution matches the real data, with lower values indicating closer similarity. As with correlation, the feature distributions for individual performances do not have zero divergence with all score interpretations, showing the diversity of musical interpretations. Of all the models, the \texttt{MLM} shows less divergence in the feature distribution from the real data. It is closely followed by the \texttt{EncDec} model, another approach that generates tokenized features. All other approaches model the real-valued features and can produce performances that deviate from the usual distributions. Compared to the external models, \texttt{PianoFlow} shows less divergence from real performances than \texttt{VirtuosoNet} and \texttt{DExter}, indicating that the model produces dynamics and tempo countours closer to the expected human expressiveness. As for the score tokens $c_s$, they improve the modeling of dynamics, timing, and articulation for the \texttt{PianoFlow} model.

\subsection{Listening Test}

\ifpdf
\begin{figure}[!t]
	\centering
	\includegraphics[width=1.\columnwidth]{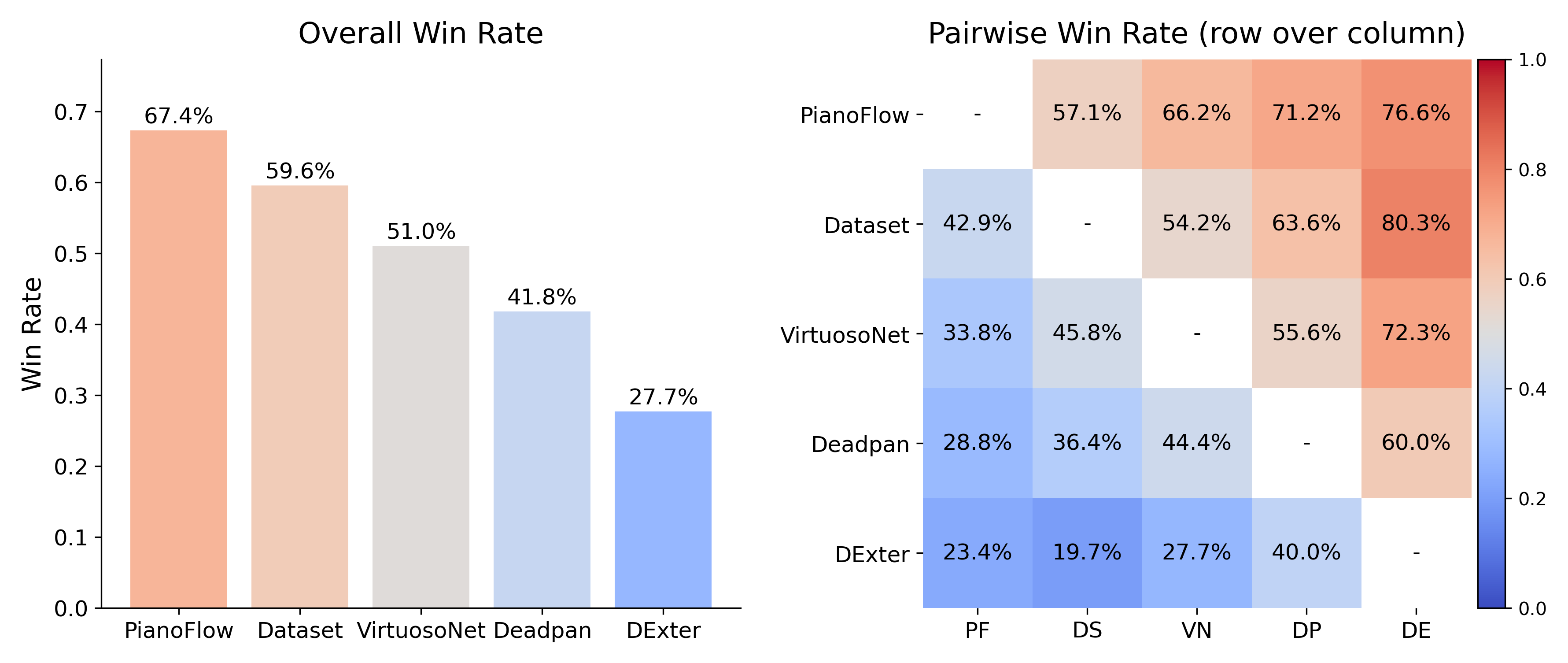}
	\caption{Results of the side-by-side listening test.}
	\label{fig:listening-test}
\end{figure}
\else\fi

In our initial internal evaluation, we compared \texttt{PianoFlow} to the \texttt{MLM} and \texttt{EncDec} baselines in a blind, side-by-side setup, with each model receiving 60 ratings. \texttt{PianoFlow} achieved a win rate of 66.7\%, \texttt{EncDec} was preferred in 57.6\% of the comparisons, while \texttt{MLM} won in 25.7\%. These results indicate that objective evaluation based on KL divergence does not reflect perceptual quality. Although \texttt{MLM} produces distributions closer to real performance features, its parallel sampling results in temporal incoherence and less pleasant performances compared to the note-by-note (\texttt{EncDec}) and the iterative, real-valued sampling (\texttt{PianoFlow}).

Next, we extended the evaluation to external participants, comparing \texttt{PianoFlow} to real data and external models. A total of 32 people with diverse musical backgrounds participated in the survey. We kept the votes of 26 participants who completed at least 10 evaluations without quick click-through responses. This phase generated 708 paired ratings, with each model, each model pair, and each composition receiving about 280, 70, and 35 votes, respectively.

Figure~\ref{fig:listening-test} summarizes the survey results. \texttt{PianoFlow} achieves the highest overall win rate of 67.1\% in pairwise comparisons. Notably, the model’s samples were even preferred over dataset samples. Two factors may explain this outcome. First, two-thirds of the real performance samples were transcribed from audio rather than recorded as MIDI. Second, listeners, who were not experienced musicians, might have found some subtle human expression nuances unfamiliar, perceiving the smoother predictions of models as more pleasing. A fair statement would be that \texttt{PianoFlow} outperforms the average combination of human MIDI performances in ASAP and ATEPP. For reference, we provide all samples on the demo web page.

\ifpdf\else
\begin{figure}[!t]
	\centering
	\includegraphics[width=1.\columnwidth]{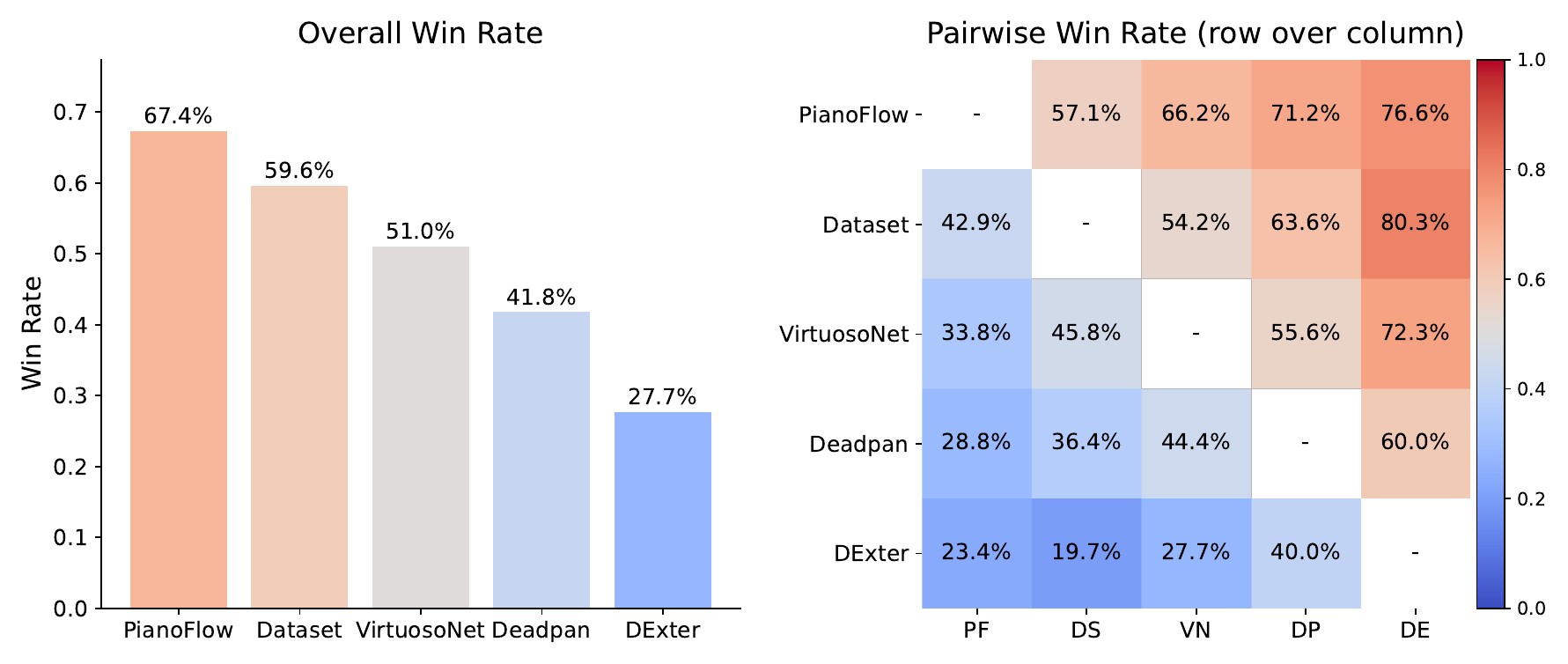}
	\caption{Results of the side-by-side listening test.}
	\label{fig:listening-test}
\end{figure}
\fi

As with the other models, \texttt{VirtuosoNet} scored slightly lower than the dataset performances but higher than deadpan performances and \texttt{DExter}. The pairwise comparisons with human data are consistent with results reported in the original work \cite{jeong2019graph} and subsequent studies \cite{worrall2024comparative, zhang2024dexter}. In contrast, the performance of \texttt{DExter} using the released checkpoint was disconcerting. One factor is that the inference code splits score in parts and synthesizes the entire performance in a batch due to a slow 1000-step reverse diffusion process.
Also, the model produced abnormal timing changes on some tested scores. The synthesized \texttt{Deadpan} scores were less preferred than the generated samples, suggesting that the models are indeed learning and reproducing expressive aspects of piano performance.

Individual participant feedback revealed that in many cases it was \textit{``easy to choose the best sample''}, while in other cases \textit{``differences were so subtle that the choices seemed almost random''}. Some also chose samples based solely on tempo and its divergence from the expectation. 
In summary, the listening test showed that \texttt{PianoFlow} captures the nuanced expressiveness of piano performance better than the baselines and is close to real performance samples.

\subsection{Emotion and Text Control}

\ifpdf
\begin{figure}[t]
	\centering
	\includegraphics[alt={},width=1.\columnwidth]{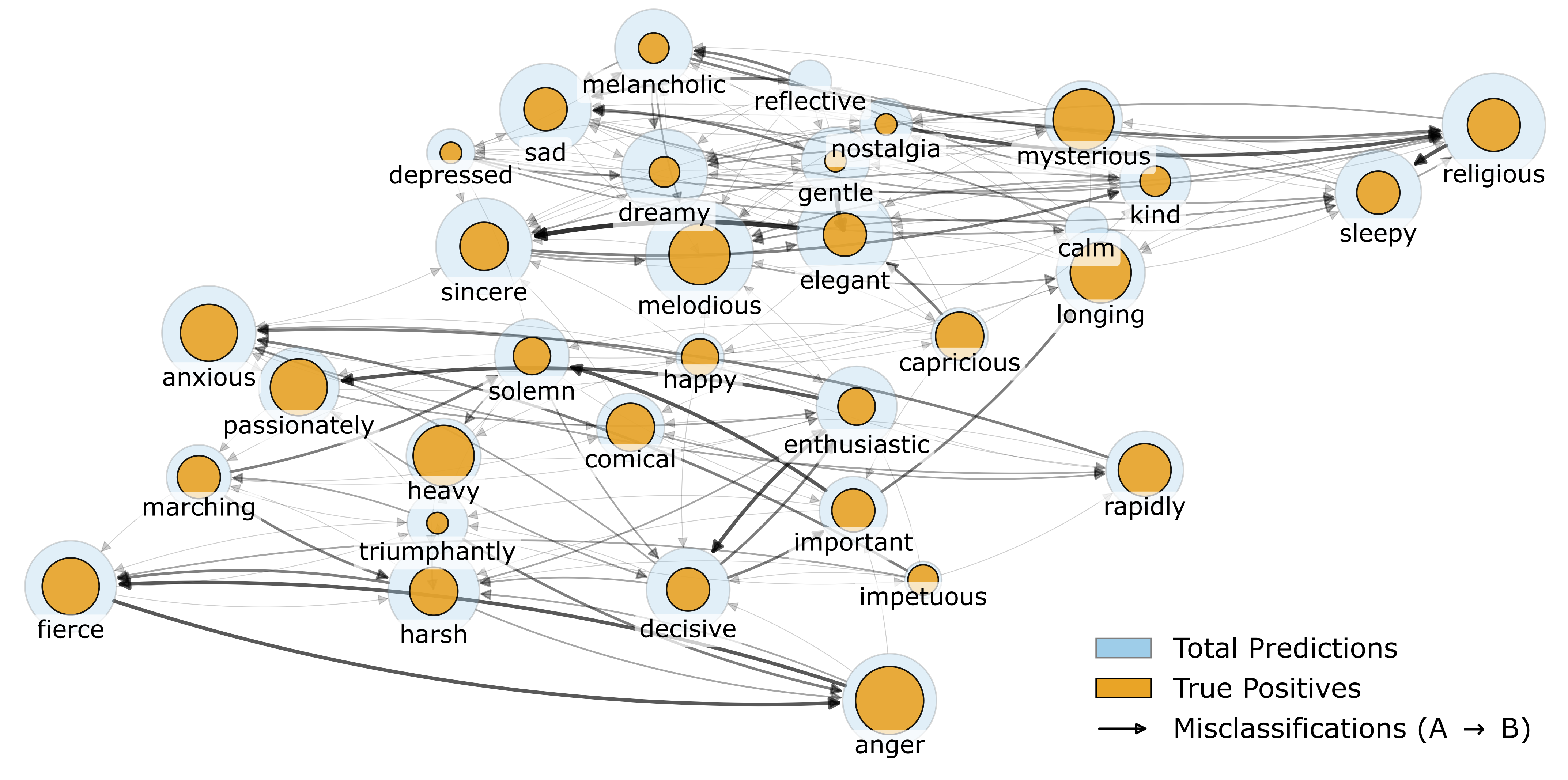}
	\caption{Emotion classifier predictions and errors represented using an emotion co-occurrence graph.}
	\label{fig:emotion-graph}
\end{figure}
\else\fi

After evaluating the unconditional generation, we study the emotion and textual control capabilities of \texttt{PianoFlow}. The trained 33 emotion piano performance classifier achieves top-1, top-3, and top-5 accuracies of 32\%, 57\%, and 73\%, respectively, on the held-out test set. Figure~\ref{fig:emotion-graph} shows the predictions errors arranged as a directed graph. The classifier groups similar-meaning emotions into clusters (e.g., loud and quiet emotions), and the errors are low-level rather than high-level. These results highlight the challenge of classifying a large set of similar emotion labels on limited data. Since emotion classification is not our primary focus, we only study the emotion- and text-conditioned inference of the PianoFlow model, which was fine-tuned with softly classified performed musical bars. 

\ifpdf\else
\begin{figure}[t]
	\centering
	\includegraphics[alt={},width=1.\columnwidth]{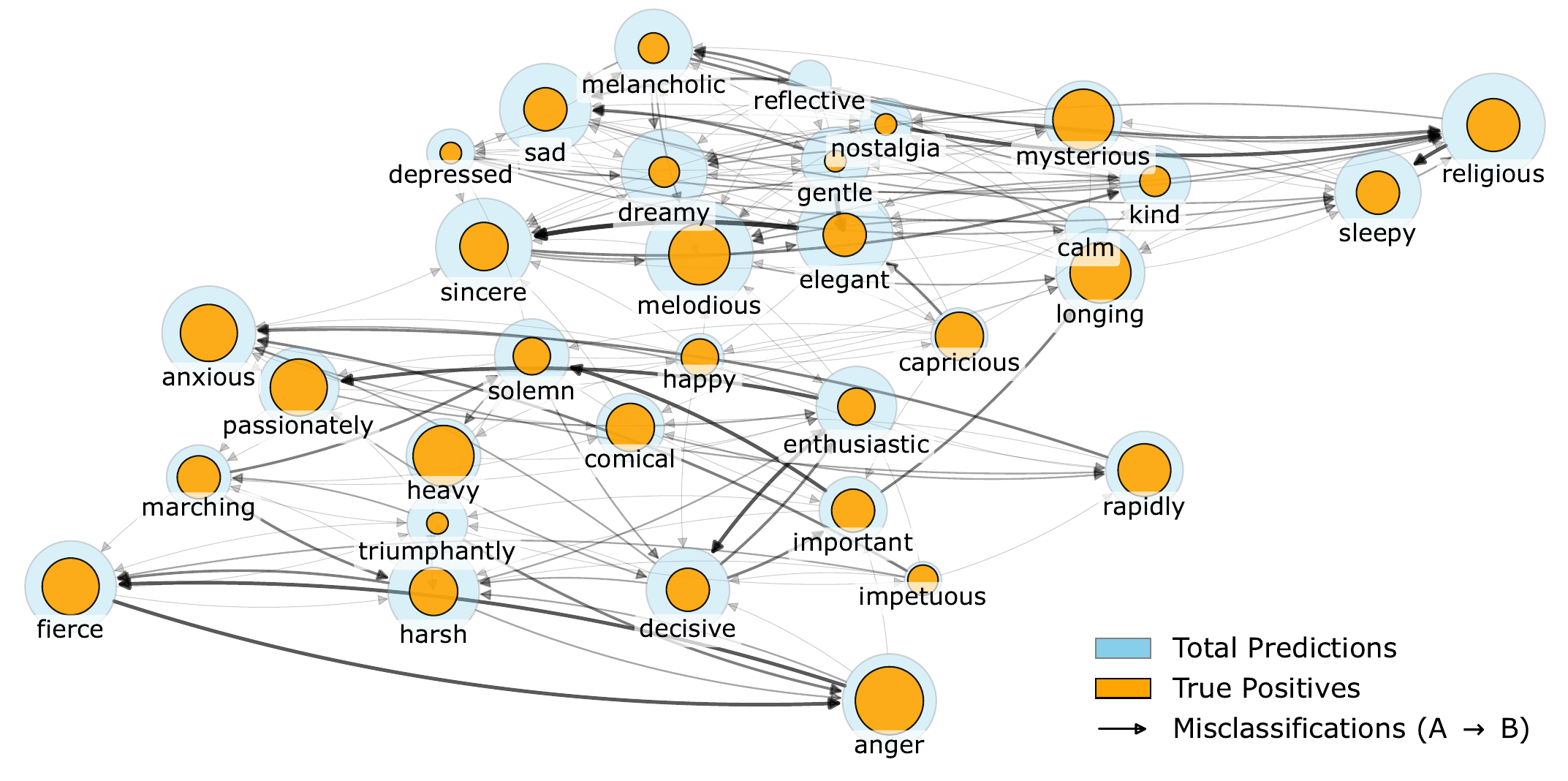}
	\caption{Emotion classifier predictions and errors represented using an emotion co-occurrence graph.}
	\label{fig:emotion-graph}
\end{figure}
\fi

\ifpdf
\begin{figure*}[t]
	\centering
	\includegraphics[alt={},width=1.\textwidth]{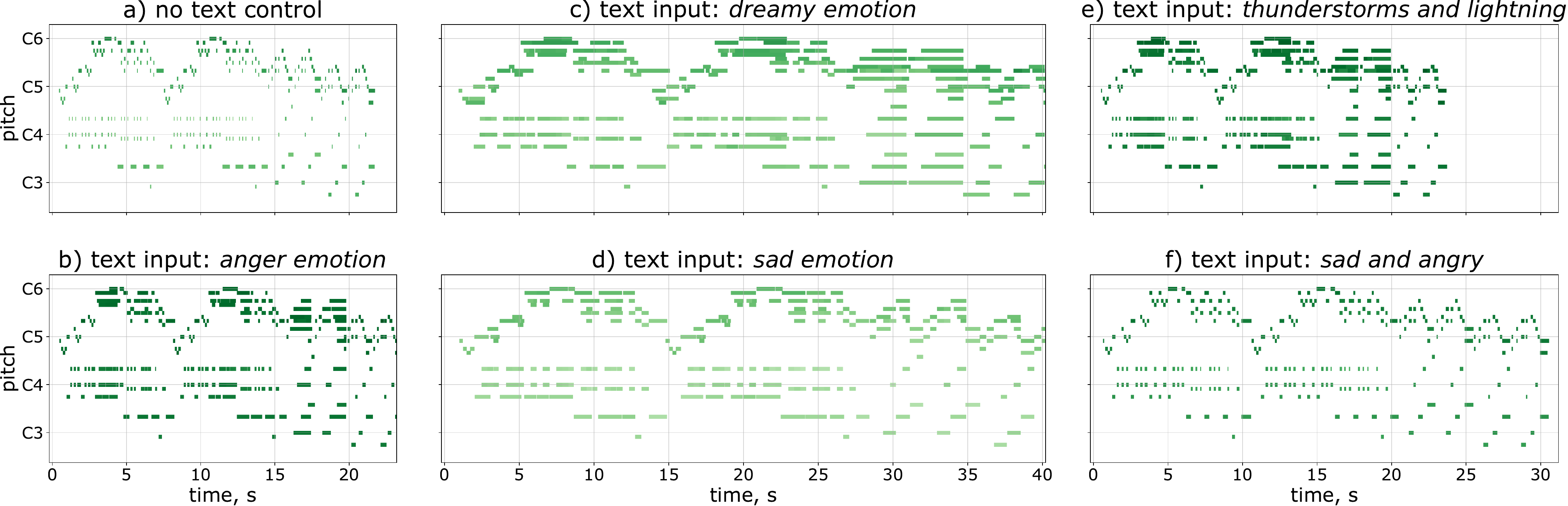}
	\caption{Emotion and text-controlled inference for the first bars of the \textit{Mozart's Piano Sonata No. 11, Mov. 3 ``Alla Turca.''}}
	\label{fig:emotion-control}
\end{figure*}

\begin{figure}[t]
	\centering
	\includegraphics[alt={},width=1.\columnwidth]{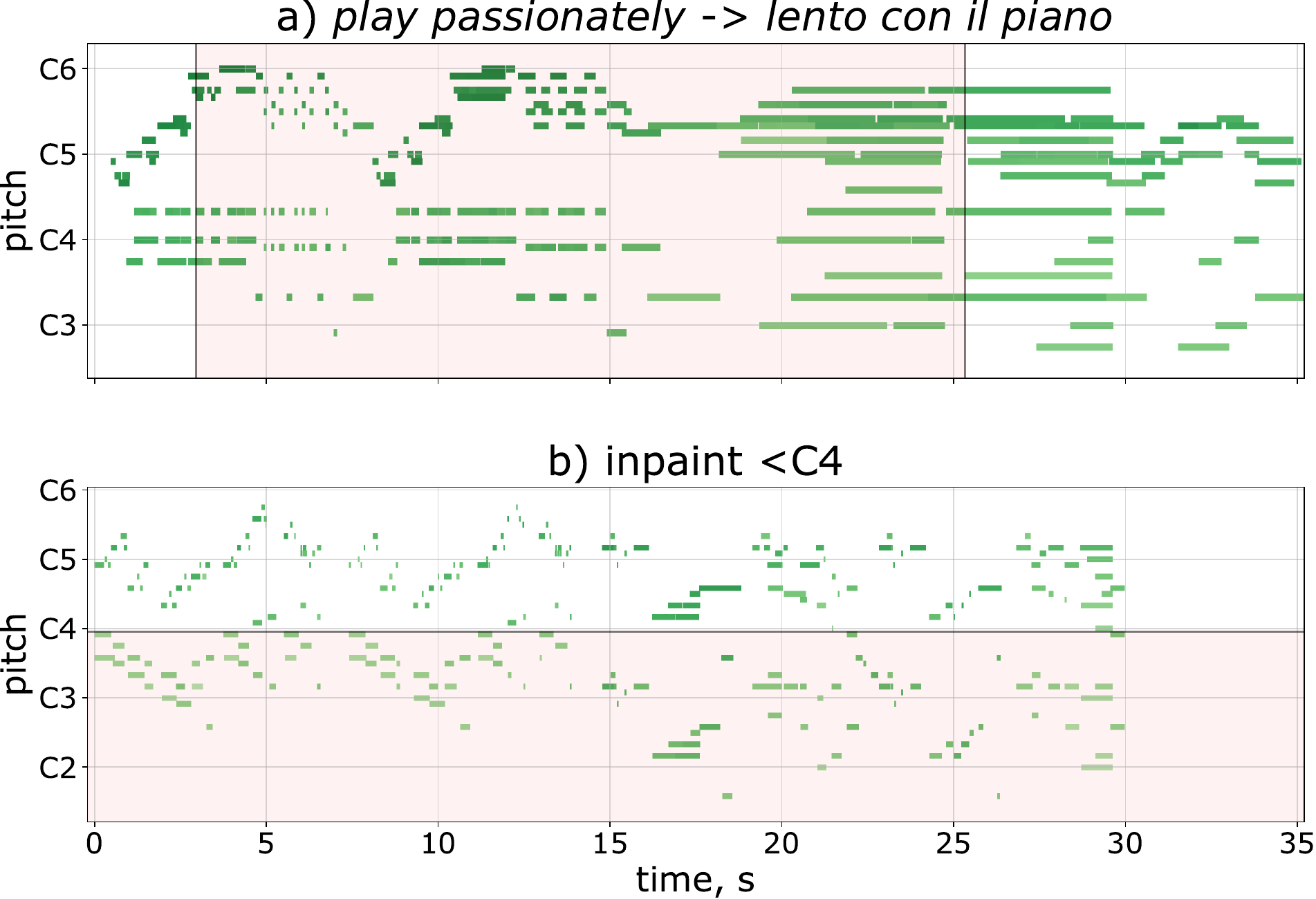}
	\caption{Inference with inpainting. The notes in the red regions were performed by the model.}
	\label{fig:inpainting}
\end{figure}
\else\fi

Figure~\ref{fig:emotion-control} shows a glimpse of what is possible with the text- and emotion-driven PianoFlow model. For testing, we take the first bars of \textit{Mozart's Piano Sonata No. 11, Movement 3 ``Alla Turca''}. The first two subplot columns visualize the unconditional performance rendering and inference using the emotions available in the data. For the emotion \textit{``anger''} (b), the performance becomes louder (brighter note color) and the notes are held with more sustain. In the middle column we observe the slower performances as a result of applying the \textit{``dreamy''} (c) and \textit{``sad''} (d) emotion control, which correlates with the expectations. Finally, the last column shows the out-of-scope text-based control. The text \textit{``thunderstorms and lightning''} (e) encoded in the embedding space of Flan-T5 provides a louder and more striking performance. The tempo is about the same, which means that the \textit{``lightning''} and fast tempo emotions are not close in the text space. And the last example (f) shows the combination of two different emotions in one text: \textit{``anger''} and \textit{``sad''}. The resulting performance is something in between: louder than just sad and slower than just angry. 
Overall, these results indicate the potential of the text-controlled expressive performance rendering.

\ifpdf\else
\begin{figure*}[t]
	\centering
	\includegraphics[alt={},width=1.\textwidth]{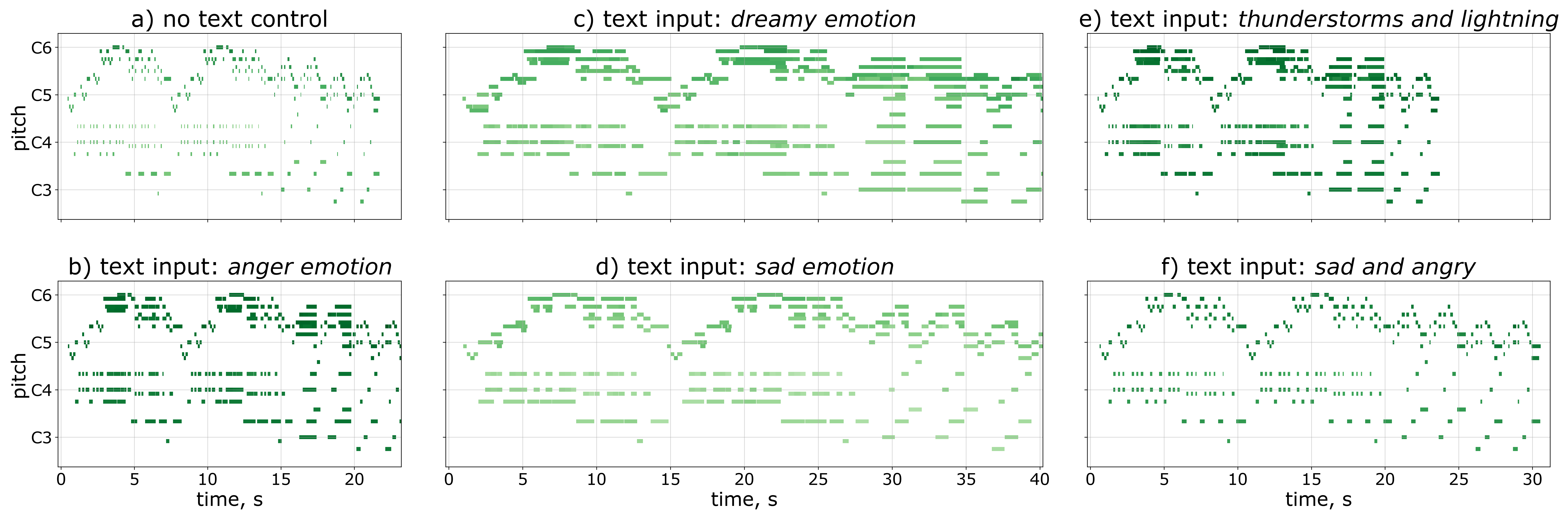}
	\caption{Emotion and text-controlled inference for the first bars of the \textit{Mozart's Piano Sonata No. 11, Mov. 3 ``Alla Turca.''}}
	\label{fig:emotion-control}
\end{figure*}
\fi

\subsection{Performance Inpainting}

As described in Section~\ref{subsec:training}, PianoFlow has been trained to solve various multi-mask performance inpainting tasks. Figure~\ref{fig:inpainting} presents two examples of universal performance inpainting. The first shows the transition between two opposite performance styles on the same \textit{Mozart's ``Alla Turca''}: from \textit{``passionate''} to \textit{``slow and quiet''}. We first generated both styles using text prompts, kept the first 32 notes of one and the last 32 of the other, and let the model fill in the middle. While the transition isn't perfectly smooth, the style changes gradually over time and align with musical bar boundaries. Despite being a challenging case, the example demonstrates the model’s ability to shift between styles fluidly.
In the second example, we mask the left-hand notes from the beginning of \textit{Beethoven’s Piano Sonata No. 25, 3rd movement}. The model effectively complements the right-hand expression, producing a full-sounding performance. 
Inpainting can also be used to recover individual notes, for example, those removed or interpolated during alignment cleanup, or to restore missing sections, such as a skipped repeat, while preserving the original style.
% Audio samples are available on the demo page.

\ifpdf\else
\begin{figure}[t]
	\centering
	\includegraphics[alt={},width=1.\columnwidth]{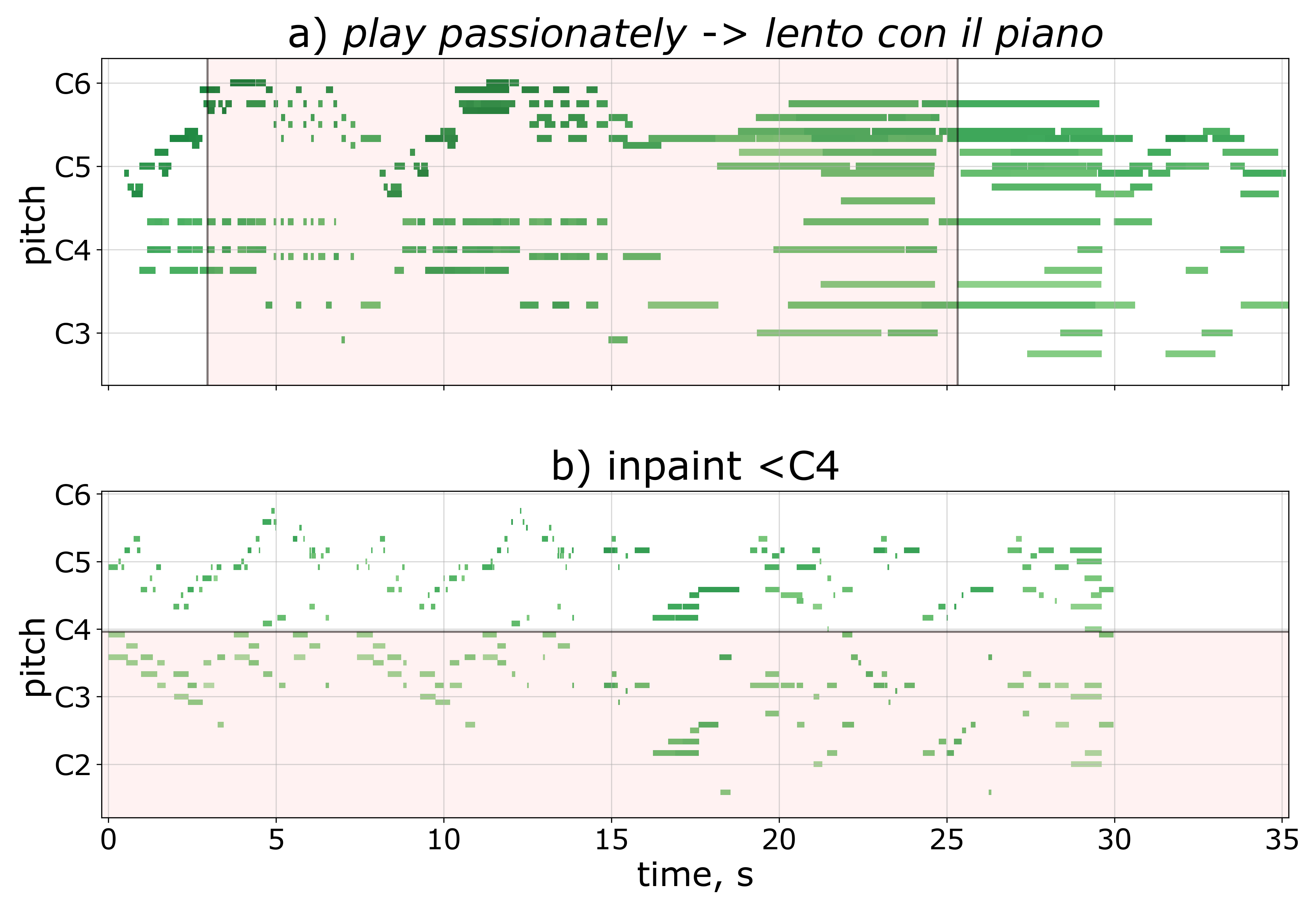}
	\caption{Inference with inpainting. The notes in the red regions were performed by the model.}
	\label{fig:inpainting}
\end{figure}
\fi

\section{Conclusion}

We presented SyMuPe, a framework for creating and training transformer models for rendering expressive music performances. Within this framework, we developed and compared three models: a masked language modeling encoder, a causal language modeling encoder-decoder, and a transformer using conditional flow matching objective. The latter model, PianoFlow, achieves state-of-the-art performance in unconditional piano performance rendering, outperforming existing solutions. In particular, in a blind, side-by-side listening test, PianoFlow was preferred over real MIDI performances, demonstrating its ability to capture nuanced expressiveness. Moreover, by integrating arbitrary textual and emotional inputs, PianoFlow enables intuitive, multimodal control of performance rendering, enhancing both interpretability and user experience. With its support for real-time inference, our approach can be easily embedded in interactive music performance systems. SyMuPe and PianoFlow represent a significant step toward universal, simple, and effective models for expressive music performance rendering.

\section*{Acknowledgements}\label{sec:acknowledgements}

The authors acknowledge the use of
Zhores supercomputer \cite{zacharov2019zhores} for obtaining the results presented in this manuscript. The authors thank the participants of the listening test for dedicating their time in evaluating the quality of the models. The authors thank the reviewers for their positive and valuable feedback, which helped to improve the quality of this work.

\bibliographystyle{ACM-Reference-Format}
\balance
\bibliography{references}


\begin{thebibliography}{64}

%%% the Web address.

\ifx \showCODEN    \undefined \def \showCODEN     #1{\unskip}     \fi
\ifx \showISBNx    \undefined \def \showISBNx     #1{\unskip}     \fi
\ifx \showISBNxiii \undefined \def \showISBNxiii  #1{\unskip}     \fi
\ifx \showISSN     \undefined \def \showISSN      #1{\unskip}     \fi
\ifx \showLCCN     \undefined \def \showLCCN      #1{\unskip}     \fi
\ifx \shownote     \undefined \def \shownote      #1{#1}          \fi
\ifx \showarticletitle \undefined \def \showarticletitle #1{#1}   \fi
\ifx \showURL      \undefined \def \showURL       {\relax}        \fi
% invisible to TeX
\providecommand\bibfield[2]{#2}
\providecommand\bibinfo[2]{#2}
\providecommand\natexlab[1]{#1}
\providecommand\showeprint[2][]{arXiv:#2}

\bibitem[Bao and Sun(2022)]%
        {bao2022generating}
\bibfield{author}{\bibinfo{person}{Chunhui Bao} {and} \bibinfo{person}{Qianru Sun}.} \bibinfo{year}{2022}\natexlab{}.
\newblock \showarticletitle{{Generating Music With Emotions}}.
\newblock \bibinfo{journal}{\emph{IEEE Transactions on Multimedia}}  \bibinfo{volume}{25} (\bibinfo{year}{2022}), \bibinfo{pages}{3602--3614}.
\newblock
\href{https://doi.org/10.1109/TMM.2022.3163543}{doi:\nolinkurl{10.1109/TMM.2022.3163543}}


\bibitem[Borovik(2021)]%
        {borovik2021thesis}
\bibfield{author}{\bibinfo{person}{Ilya Borovik}.} \bibinfo{year}{2021}\natexlab{}.
\newblock \emph{\bibinfo{title}{{Modeling Expressive Music Performance with Transformers}}}.
\newblock \bibinfo{thesistype}{Master's\ thesis}. \bibinfo{school}{Skolkovo Institute of Science and Technology}, \bibinfo{address}{Russia}.
\newblock


\bibitem[Borovik and Viro(2023a)]%
        {borovik2023hhai}
\bibfield{author}{\bibinfo{person}{Ilya Borovik} {and} \bibinfo{person}{Vladimir Viro}.} \bibinfo{year}{2023}\natexlab{a}.
\newblock \showarticletitle{{Co-Performing Music with {AI}: Real-Time Performance Control Using Speech and Gestures}}.
\newblock In \bibinfo{booktitle}{\emph{HHAI 2023: Augmenting Human Intellect}}. \bibinfo{publisher}{IOS Press}, \bibinfo{pages}{340--350}.
\newblock
\href{https://doi.org/10.3233/FAIA230097}{doi:\nolinkurl{10.3233/FAIA230097}}


\bibitem[Borovik and Viro(2023b)]%
        {borovik2023nime}
\bibfield{author}{\bibinfo{person}{Ilya Borovik} {and} \bibinfo{person}{Vladimir Viro}.} \bibinfo{year}{2023}\natexlab{b}.
\newblock \showarticletitle{Real-Time Co-Creation of Expressive Music Performances Using Speech and Gestures}. In \bibinfo{booktitle}{\emph{Proceedings of the 23rd International Conference on New Interfaces for Musical Expression (NIME)}}. \bibinfo{pages}{620--625}.
\newblock
\showISSN{2220-4806}
\href{https://doi.org/10.5281/zenodo.11189321}{doi:\nolinkurl{10.5281/zenodo.11189321}}


\bibitem[Borovik and Viro(2023c)]%
        {borovik2023scoreperformer}
\bibfield{author}{\bibinfo{person}{Ilya Borovik} {and} \bibinfo{person}{Vladimir Viro}.} \bibinfo{year}{2023}\natexlab{c}.
\newblock \showarticletitle{{ScorePerformer: Expressive Piano Performance Rendering with Fine-Grained Control}}. In \bibinfo{booktitle}{\emph{Proceedings of the 24th International Society for Music Information Retrieval Conference {(ISMIR)}}}. \bibinfo{pages}{588--596}.
\newblock
\href{https://doi.org/10.5281/zenodo.10265355}{doi:\nolinkurl{10.5281/zenodo.10265355}}


\bibitem[Cancino-Chac{\'o}n(2018)]%
        {cancino2018thesis}
\bibfield{author}{\bibinfo{person}{Carlos~Eduardo Cancino-Chac{\'o}n}.} \bibinfo{year}{2018}\natexlab{}.
\newblock \emph{\bibinfo{title}{{Computational Modeling of Expressive Music Performance with Linear and Non-linear Basis Function Models}}}.
\newblock \bibinfo{thesistype}{Ph.\,D. Dissertation}. \bibinfo{school}{Johannes Kepler University Linz}, \bibinfo{address}{Austria}.
\newblock


\bibitem[Cancino-Chac{\'o}n et~al\mbox{.}(2018)]%
        {cancino2018computational}
\bibfield{author}{\bibinfo{person}{Carlos~Eduardo Cancino-Chac{\'o}n}, \bibinfo{person}{Maarten Grachten}, \bibinfo{person}{Werner Goebl}, {and} \bibinfo{person}{Gerhard Widmer}.} \bibinfo{year}{2018}\natexlab{}.
\newblock \showarticletitle{{Computational Models of Expressive Music Performance: A Comprehensive and Critical Review}}.
\newblock \bibinfo{journal}{\emph{Frontiers in Digital Humanities}}  \bibinfo{volume}{5} (\bibinfo{year}{2018}), \bibinfo{pages}{25}.
\newblock
\href{https://doi.org/10.3389/fdigh.2018.00025}{doi:\nolinkurl{10.3389/fdigh.2018.00025}}


\bibitem[Cancino-Chac{\'o}n et~al\mbox{.}(2022)]%
        {cancino2022partitura}
\bibfield{author}{\bibinfo{person}{Carlos~Eduardo Cancino-Chac{\'o}n}, \bibinfo{person}{Silvan~David Peter}, \bibinfo{person}{Emmanouil Karystinaios}, \bibinfo{person}{Francesco Foscarin}, \bibinfo{person}{Maarten Grachten}, {and} \bibinfo{person}{Gerhard Widmer}.} \bibinfo{year}{2022}\natexlab{}.
\newblock \showarticletitle{{Partitura: A Python Package for Symbolic Music Processing}}. In \bibinfo{booktitle}{\emph{{Proceedings of the Music Encoding Conference (MEC)}}}. \bibinfo{address}{Halifax, Canada}.
\newblock
\href{https://doi.org/10.17613/131v-k502}{doi:\nolinkurl{10.17613/131v-k502}}


\bibitem[Chen et~al\mbox{.}(2024)]%
        {chen2024f5}
\bibfield{author}{\bibinfo{person}{Yushen Chen}, \bibinfo{person}{Zhikang Niu}, \bibinfo{person}{Ziyang Ma}, \bibinfo{person}{Keqi Deng}, \bibinfo{person}{Chunhui Wang}, \bibinfo{person}{Jian Zhao}, \bibinfo{person}{Kai Yu}, {and} \bibinfo{person}{Xie Chen}.} \bibinfo{year}{2024}\natexlab{}.
\newblock \showarticletitle{{F5-TTS: A Fairytaler that Fakes Fluent and Faithful Speech with Flow Matching}}.
\newblock \bibinfo{journal}{\emph{arXiv preprint arXiv:2410.06885}} (\bibinfo{year}{2024}).
\newblock
\href{https://doi.org/10.48550/arXiv.2410.06885}{doi:\nolinkurl{10.48550/arXiv.2410.06885}}


\bibitem[Chung et~al\mbox{.}(2024)]%
        {chung2024scaling}
\bibfield{author}{\bibinfo{person}{Hyung~Won Chung}, \bibinfo{person}{Le Hou}, \bibinfo{person}{Shayne Longpre}, \bibinfo{person}{Barret Zoph}, \bibinfo{person}{Yi Tay}, \bibinfo{person}{William Fedus}, \bibinfo{person}{Yunxuan Li}, \bibinfo{person}{Xuezhi Wang}, \bibinfo{person}{Mostafa Dehghani}, \bibinfo{person}{Siddhartha Brahma}, {et~al\mbox{.}}} \bibinfo{year}{2024}\natexlab{}.
\newblock \showarticletitle{{Scaling Instruction-Finetuned Language Models}}.
\newblock \bibinfo{journal}{\emph{Journal of Machine Learning Research}} \bibinfo{volume}{25}, \bibinfo{number}{70} (\bibinfo{year}{2024}), \bibinfo{pages}{1--53}.
\newblock


\bibitem[Dash and Agres(2024)]%
        {dash2024affective}
\bibfield{author}{\bibinfo{person}{Adyasha Dash} {and} \bibinfo{person}{Kathleen Agres}.} \bibinfo{year}{2024}\natexlab{}.
\newblock \showarticletitle{{AI-Based Affective Music Generation Systems: A Review of Methods and Challenges}}.
\newblock \bibinfo{journal}{\emph{Comput. Surveys}} \bibinfo{volume}{56}, \bibinfo{number}{11} (\bibinfo{year}{2024}), \bibinfo{pages}{1--34}.
\newblock
\href{https://doi.org/10.1145/3672554}{doi:\nolinkurl{10.1145/3672554}}


\bibitem[Fei et~al\mbox{.}(2024)]%
        {fei2024flux}
\bibfield{author}{\bibinfo{person}{Zhengcong Fei}, \bibinfo{person}{Mingyuan Fan}, \bibinfo{person}{Changqian Yu}, {and} \bibinfo{person}{Junshi Huang}.} \bibinfo{year}{2024}\natexlab{}.
\newblock \showarticletitle{{FLUX that Plays Music}}.
\newblock \bibinfo{journal}{\emph{arXiv preprint arXiv:2409.00587}} (\bibinfo{year}{2024}).
\newblock
\href{https://doi.org/10.48550/arXiv.2409.00587}{doi:\nolinkurl{10.48550/arXiv.2409.00587}}


\bibitem[Foscarin et~al\mbox{.}(2020)]%
        {foscarin2020asap}
\bibfield{author}{\bibinfo{person}{Francesco Foscarin}, \bibinfo{person}{Andrew Mcleod}, \bibinfo{person}{Philippe Rigaux}, \bibinfo{person}{Florent Jacquemard}, {and} \bibinfo{person}{Masahiko Sakai}.} \bibinfo{year}{2020}\natexlab{}.
\newblock \showarticletitle{{ASAP: A Dataset of Aligned Scores and Performances for Piano Transcription}}. In \bibinfo{booktitle}{\emph{Proceedings of the 21st International Society for Music Information Retrieval Conference {(ISMIR)}}}. \bibinfo{pages}{534--541}.
\newblock
\href{https://doi.org/10.5281/zenodo.4245490}{doi:\nolinkurl{10.5281/zenodo.4245490}}


\bibitem[Fradet et~al\mbox{.}(2021)]%
        {fradet2021miditok}
\bibfield{author}{\bibinfo{person}{Nathan Fradet}, \bibinfo{person}{Jean-Pierre Briot}, \bibinfo{person}{Fabien Chhel}, \bibinfo{person}{Amal El~Fallah~Seghrouchni}, {and} \bibinfo{person}{Nicolas Gutowski}.} \bibinfo{year}{2021}\natexlab{}.
\newblock \showarticletitle{{MidiTok}: A Python package for {MIDI} file tokenization}. In \bibinfo{booktitle}{\emph{Extended Abstracts for the Late-Breaking Demo Session of the 22nd International Society for Music Information Retrieval Conference {(ISMIR)}}}.
\newblock


\bibitem[Gabrielsson and Juslin(1996)]%
        {gabrielsson1996emotional}
\bibfield{author}{\bibinfo{person}{Alf Gabrielsson} {and} \bibinfo{person}{Patrik~N. Juslin}.} \bibinfo{year}{1996}\natexlab{}.
\newblock \showarticletitle{Emotional expression in music performance: Between the performer's intention and the listener's experience}.
\newblock \bibinfo{journal}{\emph{Psychology of music}} \bibinfo{volume}{24}, \bibinfo{number}{1} (\bibinfo{year}{1996}), \bibinfo{pages}{68--91}.
\newblock
\href{https://doi.org/10.1177/0305735696241007}{doi:\nolinkurl{10.1177/0305735696241007}}


\bibitem[Good(2001)]%
        {good2001musicxml}
\bibfield{author}{\bibinfo{person}{Michael Good}.} \bibinfo{year}{2001}\natexlab{}.
\newblock \showarticletitle{{MusicXML for Notation and Analysis }}.
\newblock \bibinfo{journal}{\emph{The Virtual Score: Representation, Retrieval, Restoration}} \bibinfo{volume}{12}, \bibinfo{number}{113-124} (\bibinfo{year}{2001}), \bibinfo{pages}{160}.
\newblock
\href{https://doi.org/10.7551/mitpress/2058.003.0010}{doi:\nolinkurl{10.7551/mitpress/2058.003.0010}}


\bibitem[Guo et~al\mbox{.}(2024)]%
        {guo2024voiceflow}
\bibfield{author}{\bibinfo{person}{Yiwei Guo}, \bibinfo{person}{Chenpeng Du}, \bibinfo{person}{Ziyang Ma}, \bibinfo{person}{Xie Chen}, {and} \bibinfo{person}{Kai Yu}.} \bibinfo{year}{2024}\natexlab{}.
\newblock \showarticletitle{{Voiceflow: Efficient text-to-speech with rectified flow matching}}. In \bibinfo{booktitle}{\emph{Proceeding of the 49th IEEE International Conference on Acoustics, Speech and Signal Processing (ICASSP)}}. IEEE, \bibinfo{pages}{11121--11125}.
\newblock
\href{https://doi.org/10.1109/ICASSP48485.2024.10445948}{doi:\nolinkurl{10.1109/ICASSP48485.2024.10445948}}


\bibitem[Guo et~al\mbox{.}(2023)]%
        {guo2023fme}
\bibfield{author}{\bibinfo{person}{Zixun Guo}, \bibinfo{person}{Jaeyong Kang}, {and} \bibinfo{person}{Dorien Herremans}.} \bibinfo{year}{2023}\natexlab{}.
\newblock \showarticletitle{{A Domain-Knowledge-Inspired Music Embedding Space and a Novel Attention Mechanism for Symbolic Music Modeling}}. In \bibinfo{booktitle}{\emph{Proceedings of the 37th AAAI Conference on Artificial Intelligence}}, Vol.~\bibinfo{volume}{37}. \bibinfo{pages}{5070--5077}.
\newblock
\href{https://doi.org/10.1609/aaai.v37i4.25635}{doi:\nolinkurl{10.1609/aaai.v37i4.25635}}


\bibitem[Ho and Salimans(2021)]%
        {ho2022cfg}
\bibfield{author}{\bibinfo{person}{Jonathan Ho} {and} \bibinfo{person}{Tim Salimans}.} \bibinfo{year}{2021}\natexlab{}.
\newblock \showarticletitle{{Classifier-Free Diffusion Guidance}}. In \bibinfo{booktitle}{\emph{NeurIPS 2021 Workshop on Deep Generative Models and Downstream Applications}}.
\newblock


\bibitem[Hsiao et~al\mbox{.}(2021)]%
        {hsiao2021compound}
\bibfield{author}{\bibinfo{person}{Wen-Yi Hsiao}, \bibinfo{person}{Jen-Yu Liu}, \bibinfo{person}{Yin-Cheng Yeh}, {and} \bibinfo{person}{Yi-Hsuan Yang}.} \bibinfo{year}{2021}\natexlab{}.
\newblock \showarticletitle{{Compound Word Transformer: Learning to Compose Full-Song Music over Dynamic Directed Hypergraphs}}. In \bibinfo{booktitle}{\emph{Proceedings of the 35th AAAI Conference on Artificial Intelligence}}, Vol.~\bibinfo{volume}{35}. \bibinfo{pages}{178--186}.
\newblock
\href{https://doi.org/10.1609/aaai.v35i1.16091}{doi:\nolinkurl{10.1609/aaai.v35i1.16091}}


\bibitem[Huang et~al\mbox{.}(2024)]%
        {huang2024emotion}
\bibfield{author}{\bibinfo{person}{Jingyue Huang}, \bibinfo{person}{Ke Chen}, {and} \bibinfo{person}{Yi-Hsuan Yang}.} \bibinfo{year}{2024}\natexlab{}.
\newblock \showarticletitle{{Emotion-driven Piano Music Generation via Two-stage Disentanglement and Functional Representation}}. In \bibinfo{booktitle}{\emph{Proceedings of the 25th International Society for Music Information Retrieval Conference {(ISMIR)}}}.
\newblock
\href{https://doi.org/10.5281/zenodo.14877343}{doi:\nolinkurl{10.5281/zenodo.14877343}}


\bibitem[Huang and Yang(2020)]%
        {huang2020pop}
\bibfield{author}{\bibinfo{person}{Yu-Siang Huang} {and} \bibinfo{person}{Yi-Hsuan Yang}.} \bibinfo{year}{2020}\natexlab{}.
\newblock \showarticletitle{{Pop Music Transformer: Beat-based Modeling and Generation of Expressive Pop Piano Compositions}}. In \bibinfo{booktitle}{\emph{Proceedings of the 28th ACM International Conference on Multimedia}}. \bibinfo{pages}{1180--1188}.
\newblock
\href{https://doi.org/10.1145/3394171.3413671}{doi:\nolinkurl{10.1145/3394171.3413671}}


\bibitem[Hung et~al\mbox{.}(2021)]%
        {hung2021emopia}
\bibfield{author}{\bibinfo{person}{Hsiao-Tzu Hung}, \bibinfo{person}{Joann Ching}, \bibinfo{person}{Seungheon Doh}, \bibinfo{person}{Nabin Kim}, \bibinfo{person}{Juhan Nam}, {and} \bibinfo{person}{Yi-Hsuan Yang}.} \bibinfo{year}{2021}\natexlab{}.
\newblock \showarticletitle{{EMOPIA: A multi-modal pop piano dataset for emotion recognition and emotion-based music generation}}. In \bibinfo{booktitle}{\emph{Proceedings of the 22nd International Society for Music Information Retrieval Conference {(ISMIR)}}}.
\newblock
\href{https://doi.org/10.5281/zenodo.5624519}{doi:\nolinkurl{10.5281/zenodo.5624519}}


\bibitem[Hung et~al\mbox{.}(2024)]%
        {hung2024eme33}
\bibfield{author}{\bibinfo{person}{Tzu-Ching Hung}, \bibinfo{person}{Jingjing Tang}, \bibinfo{person}{Kit Armstrong}, \bibinfo{person}{Yi-Cheng Lin}, {and} \bibinfo{person}{Yi-Wen Liu}.} \bibinfo{year}{2024}\natexlab{}.
\newblock \showarticletitle{EME33: A Dataset of Classical Piano Performances Guided by Expressive Markings with Application in Music Rendering}. In \bibinfo{booktitle}{\emph{2024 IEEE International Conference on Big Data (BigData)}}. IEEE, \bibinfo{pages}{3174--3180}.
\newblock
\href{https://doi.org/10.1109/BigData62323.2024.10826039}{doi:\nolinkurl{10.1109/BigData62323.2024.10826039}}


\bibitem[Jeong et~al\mbox{.}(2019c)]%
        {jeong2019virtuosonet}
\bibfield{author}{\bibinfo{person}{Dasaem Jeong}, \bibinfo{person}{Taegyun Kwon}, \bibinfo{person}{Yoojin Kim}, \bibinfo{person}{Kyogu Lee}, {and} \bibinfo{person}{Juhan Nam}.} \bibinfo{year}{2019}\natexlab{c}.
\newblock \showarticletitle{{VirtuosoNet: A Hierarchical RNN-based System for Modeling Expressive Piano Performance}}. In \bibinfo{booktitle}{\emph{Proceedings of the 20th International Society for Music Information Retrieval Conference {(ISMIR)}}}. \bibinfo{pages}{908--915}.
\newblock
\href{https://doi.org/10.5281/zenodo.3527962}{doi:\nolinkurl{10.5281/zenodo.3527962}}


\bibitem[Jeong et~al\mbox{.}(2019a)]%
        {jeong2019graph}
\bibfield{author}{\bibinfo{person}{Dasaem Jeong}, \bibinfo{person}{Taegyun Kwon}, \bibinfo{person}{Yoojin Kim}, {and} \bibinfo{person}{Juhan Nam}.} \bibinfo{year}{2019}\natexlab{a}.
\newblock \showarticletitle{{Graph Neural Network for Music Score Data and Modeling Expressive Piano Performance}}. In \bibinfo{booktitle}{\emph{Proceedings of the 36th International Conference on Machine Learning {(ICML)}}}. \bibinfo{publisher}{PMLR}, \bibinfo{pages}{3060--3070}.
\newblock


\bibitem[Jeong et~al\mbox{.}(2019b)]%
        {jeong2019score}
\bibfield{author}{\bibinfo{person}{Dasaem Jeong}, \bibinfo{person}{Taegyun Kwon}, \bibinfo{person}{Yoojin Kim}, {and} \bibinfo{person}{Juhan Nam}.} \bibinfo{year}{2019}\natexlab{b}.
\newblock \showarticletitle{{Score and performance features for rendering expressive music performances}}. In \bibinfo{booktitle}{\emph{{Proceedings of the Music Encoding Conference (MEC)}}}. Music Encoding Initiative Vienna, Austria, \bibinfo{pages}{1--6}.
\newblock


\bibitem[Jin et~al\mbox{.}(2025)]%
        {jin2025pyramidal}
\bibfield{author}{\bibinfo{person}{Yang Jin}, \bibinfo{person}{Zhicheng Sun}, \bibinfo{person}{Ningyuan Li}, \bibinfo{person}{Kun Xu}, \bibinfo{person}{Hao Jiang}, \bibinfo{person}{Nan Zhuang}, \bibinfo{person}{Quzhe Huang}, \bibinfo{person}{Yang Song}, \bibinfo{person}{Yadong MU}, {and} \bibinfo{person}{Zhouchen Lin}.} \bibinfo{year}{2025}\natexlab{}.
\newblock \showarticletitle{Pyramidal Flow Matching for Efficient Video Generative Modeling}. In \bibinfo{booktitle}{\emph{Proceedings of the 13th International Conference on Representation Learning (ICLR)}}, Vol.~\bibinfo{volume}{2025}. \bibinfo{pages}{23378--23402}.
\newblock


\bibitem[Juslin and Laukka(2003)]%
        {juslin2003communication}
\bibfield{author}{\bibinfo{person}{Patrik~N. Juslin} {and} \bibinfo{person}{Petri Laukka}.} \bibinfo{year}{2003}\natexlab{}.
\newblock \showarticletitle{{Communication of Emotions in Vocal Expression and Music Performance: Different Channels, Same Code?}}
\newblock \bibinfo{journal}{\emph{Psychological bulletin}} \bibinfo{volume}{129}, \bibinfo{number}{5} (\bibinfo{year}{2003}), \bibinfo{pages}{770--814}.
\newblock
\href{https://doi.org/10.1037/0033-2909.129.5.770}{doi:\nolinkurl{10.1037/0033-2909.129.5.770}}


\bibitem[Kingma and Ba(2015)]%
        {kingma2014adam}
\bibfield{author}{\bibinfo{person}{Diederik~P. Kingma} {and} \bibinfo{person}{Jimmy Ba}.} \bibinfo{year}{2015}\natexlab{}.
\newblock \showarticletitle{Adam: {A} Method for Stochastic Optimization}. In \bibinfo{booktitle}{\emph{Proceedings of the 3rd International Conference on Learning Representations (ICLR)}}.
\newblock


\bibitem[Labs et~al\mbox{.}(2025)]%
        {batifol2025flux}
\bibfield{author}{\bibinfo{person}{Black~Forest Labs}, \bibinfo{person}{Stephen Batifol}, \bibinfo{person}{Andreas Blattmann}, \bibinfo{person}{Frederic Boesel}, \bibinfo{person}{Saksham Consul}, \bibinfo{person}{Cyril Diagne}, \bibinfo{person}{Tim Dockhorn}, \bibinfo{person}{Jack English}, \bibinfo{person}{Zion English}, \bibinfo{person}{Patrick Esser}, {et~al\mbox{.}}} \bibinfo{year}{2025}\natexlab{}.
\newblock \showarticletitle{FLUX. 1 Kontext: Flow Matching for In-Context Image Generation and Editing in Latent Space}.
\newblock \bibinfo{journal}{\emph{arXiv preprint arXiv:2506.15742}} (\bibinfo{year}{2025}).
\newblock
\href{https://doi.org/10.48550/arXiv.2506.15742}{doi:\nolinkurl{10.48550/arXiv.2506.15742}}


\bibitem[Le et~al\mbox{.}(2024)]%
        {le2024voicebox}
\bibfield{author}{\bibinfo{person}{Matthew Le}, \bibinfo{person}{Apoorv Vyas}, \bibinfo{person}{Bowen Shi}, \bibinfo{person}{Brian Karrer}, \bibinfo{person}{Leda Sari}, \bibinfo{person}{Rashel Moritz}, \bibinfo{person}{Mary Williamson}, \bibinfo{person}{Vimal Manohar}, \bibinfo{person}{Yossi Adi}, \bibinfo{person}{Jay Mahadeokar}, {et~al\mbox{.}}} \bibinfo{year}{2024}\natexlab{}.
\newblock \showarticletitle{{Voicebox: Text-Guided Multilingual Universal Speech Generation at Scale}}.
\newblock \bibinfo{journal}{\emph{Advances in Neural Information Processing Systems {(NeurIPS)}}}  \bibinfo{volume}{36} (\bibinfo{year}{2024}).
\newblock


\bibitem[Lipman et~al\mbox{.}(2022)]%
        {lipman2022flow}
\bibfield{author}{\bibinfo{person}{Yaron Lipman}, \bibinfo{person}{Ricky~TQ Chen}, \bibinfo{person}{Heli Ben-Hamu}, \bibinfo{person}{Maximilian Nickel}, {and} \bibinfo{person}{Matthew Le}.} \bibinfo{year}{2022}\natexlab{}.
\newblock \showarticletitle{{Flow Matching for Generative Modeling}}. In \bibinfo{booktitle}{\emph{The Proceedings of the 11th International Conference on Learning Representations {(ICLR)}}}.
\newblock


\bibitem[Lipman et~al\mbox{.}(2024)]%
        {lipman2024flow}
\bibfield{author}{\bibinfo{person}{Yaron Lipman}, \bibinfo{person}{Marton Havasi}, \bibinfo{person}{Peter Holderrieth}, \bibinfo{person}{Neta Shaul}, \bibinfo{person}{Matt Le}, \bibinfo{person}{Brian Karrer}, \bibinfo{person}{Ricky~TQ Chen}, \bibinfo{person}{David Lopez-Paz}, \bibinfo{person}{Heli Ben-Hamu}, {and} \bibinfo{person}{Itai Gat}.} \bibinfo{year}{2024}\natexlab{}.
\newblock \showarticletitle{{Flow Matching Guide and Code}}.
\newblock \bibinfo{journal}{\emph{arXiv preprint arXiv:2412.06264}} (\bibinfo{year}{2024}).
\newblock
\href{https://doi.org/10.48550/arXiv.2412.06264}{doi:\nolinkurl{10.48550/arXiv.2412.06264}}


\bibitem[Long et~al\mbox{.}(2025)]%
        {long2025pdmx}
\bibfield{author}{\bibinfo{person}{Phillip Long}, \bibinfo{person}{Zachary Novack}, \bibinfo{person}{Taylor Berg-Kirkpatrick}, {and} \bibinfo{person}{Julian McAuley}.} \bibinfo{year}{2025}\natexlab{}.
\newblock \showarticletitle{{PDMX: A Large-Scale Public Domain MusicXML Dataset for Symbolic Music Processing}}. In \bibinfo{booktitle}{\emph{Proceeding of the 50th IEEE International Conference on Acoustics, Speech and Signal Processing (ICASSP)}}. IEEE, \bibinfo{pages}{1--5}.
\newblock
\href{https://doi.org/10.1109/ICASSP49660.2025.10890217}{doi:\nolinkurl{10.1109/ICASSP49660.2025.10890217}}


\bibitem[Lundqvist et~al\mbox{.}(2009)]%
        {lundqvist2009emotional}
\bibfield{author}{\bibinfo{person}{Lars-Olov Lundqvist}, \bibinfo{person}{Fredrik Carlsson}, \bibinfo{person}{Per Hilmersson}, {and} \bibinfo{person}{Patrik~N. Juslin}.} \bibinfo{year}{2009}\natexlab{}.
\newblock \showarticletitle{{Emotional Responses to Music: Experience, Expression, and Physiology}}.
\newblock \bibinfo{journal}{\emph{Psychology of Music}} \bibinfo{volume}{37}, \bibinfo{number}{1} (\bibinfo{year}{2009}), \bibinfo{pages}{61--90}.
\newblock
\href{https://doi.org/10.1177/0305735607086048}{doi:\nolinkurl{10.1177/0305735607086048}}


\bibitem[Maezawa et~al\mbox{.}(2019)]%
        {maezawa2019rendering}
\bibfield{author}{\bibinfo{person}{Akira Maezawa}, \bibinfo{person}{Kazuhiko Yamamoto}, {and} \bibinfo{person}{Takuya Fujishima}.} \bibinfo{year}{2019}\natexlab{}.
\newblock \showarticletitle{{Rendering Music Performance With Interpretation Variations Using Conditional Variational RNN}}. In \bibinfo{booktitle}{\emph{Proceedings of the 20th International Society for Music Information Retrieval Conference {(ISMIR)}}}. \bibinfo{pages}{855--861}.
\newblock
\href{https://doi.org/10.5281/zenodo.3527948}{doi:\nolinkurl{10.5281/zenodo.3527948}}


\bibitem[Mehta et~al\mbox{.}(2024)]%
        {mehta2024matcha}
\bibfield{author}{\bibinfo{person}{Shivam Mehta}, \bibinfo{person}{Ruibo Tu}, \bibinfo{person}{Jonas Beskow}, \bibinfo{person}{{\'E}va Sz{\'e}kely}, {and} \bibinfo{person}{Gustav~Eje Henter}.} \bibinfo{year}{2024}\natexlab{}.
\newblock \showarticletitle{{Matcha-TTS: A fast TTS architecture with conditional flow matching}}. In \bibinfo{booktitle}{\emph{Proceeding of the 49th IEEE International Conference on Acoustics, Speech and Signal Processing (ICASSP)}}. IEEE, \bibinfo{pages}{11341--11345}.
\newblock
\href{https://doi.org/10.1109/ICASSP48485.2024.10448291}{doi:\nolinkurl{10.1109/ICASSP48485.2024.10448291}}


\bibitem[Nakamura et~al\mbox{.}(2017)]%
        {nakamura2017alignment}
\bibfield{author}{\bibinfo{person}{Eita Nakamura}, \bibinfo{person}{Kazuyoshi Yoshii}, {and} \bibinfo{person}{Haruhiro Katayose}.} \bibinfo{year}{2017}\natexlab{}.
\newblock \showarticletitle{{Performance Error Detection and Post-Processing for Fast and Accurate Symbolic Music Alignment}}. In \bibinfo{booktitle}{\emph{Proceedings of the 18th International Society for Music Information Retrieval Conference {(ISMIR)}}}.
\newblock
\href{https://doi.org/10.5281/zenodo.1414940}{doi:\nolinkurl{10.5281/zenodo.1414940}}


\bibitem[Palmer(1997)]%
        {palmer1997music}
\bibfield{author}{\bibinfo{person}{Caroline Palmer}.} \bibinfo{year}{1997}\natexlab{}.
\newblock \showarticletitle{{Music performance}}.
\newblock \bibinfo{journal}{\emph{Annual review of psychology}} \bibinfo{volume}{48}, \bibinfo{number}{1} (\bibinfo{year}{1997}), \bibinfo{pages}{115--138}.
\newblock
\href{https://doi.org/10.1146/annurev.psych.48.1.115}{doi:\nolinkurl{10.1146/annurev.psych.48.1.115}}


\bibitem[Peter(2023)]%
        {peter2023parangonar}
\bibfield{author}{\bibinfo{person}{Silvan~David Peter}.} \bibinfo{year}{2023}\natexlab{}.
\newblock \showarticletitle{{Online Symbolic Music Alignment with Offline Reinforcement Learning}}. In \bibinfo{booktitle}{\emph{Proceedings of the 24th International Society for Music Information Retrieval Conference {(ISMIR)}}}.
\newblock
\href{https://doi.org/10.5281/zenodo.10265367}{doi:\nolinkurl{10.5281/zenodo.10265367}}


\bibitem[Peter et~al\mbox{.}(2023)]%
        {peter2023nasap}
\bibfield{author}{\bibinfo{person}{Silvan~David Peter}, \bibinfo{person}{Carlos~Eduardo Cancino-Chac{\'o}n}, \bibinfo{person}{Francesco Foscarin}, \bibinfo{person}{Andrew~Philip McLeod}, \bibinfo{person}{Florian Henkel}, \bibinfo{person}{Emmanouil Karystinaios}, {and} \bibinfo{person}{Gerhard Widmer}.} \bibinfo{year}{2023}\natexlab{}.
\newblock \showarticletitle{Automatic Note-Level Score-to-Performance Alignments in the ASAP Dataset}.
\newblock \bibinfo{journal}{\emph{Transactions of the International Society for Music Information Retrieval {(TISMIR)}}} (\bibinfo{year}{2023}).
\newblock
\href{https://doi.org/10.5334/tismir.149}{doi:\nolinkurl{10.5334/tismir.149}}


\bibitem[Prajwal et~al\mbox{.}(2024)]%
        {prajwal2024musicflow}
\bibfield{author}{\bibinfo{person}{KR Prajwal}, \bibinfo{person}{Bowen Shi}, \bibinfo{person}{Matthew Le}, \bibinfo{person}{Apoorv Vyas}, \bibinfo{person}{Andros Tjandra}, \bibinfo{person}{Mahi Luthra}, \bibinfo{person}{Baishan Guo}, \bibinfo{person}{Huiyu Wang}, \bibinfo{person}{Triantafyllos Afouras}, \bibinfo{person}{David Kant}, {et~al\mbox{.}}} \bibinfo{year}{2024}\natexlab{}.
\newblock \showarticletitle{{MusicFlow: Cascaded Flow Matching for Text Guided Music Generation}}. In \bibinfo{booktitle}{\emph{Proceedings of the 41st International Conference on Machine Learning {(ICML)}}}. PMLR, \bibinfo{pages}{41052--41063}.
\newblock


\bibitem[Renault et~al\mbox{.}(2023)]%
        {renault2023expressive}
\bibfield{author}{\bibinfo{person}{Lenny Renault}, \bibinfo{person}{R{\'e}mi Mignot}, {and} \bibinfo{person}{Axel Roebel}.} \bibinfo{year}{2023}\natexlab{}.
\newblock \showarticletitle{{Expressive Piano Performance Rendering from Unpaired Data}}. In \bibinfo{booktitle}{\emph{International Conference on Digital Audio Effects (DAFx23)}}.
\newblock
\href{https://doi.org/10.5281/zenodo.8386761}{doi:\nolinkurl{10.5281/zenodo.8386761}}


\bibitem[Rhyu et~al\mbox{.}(2022)]%
        {rhyu2022sketching}
\bibfield{author}{\bibinfo{person}{Seungyeon Rhyu}, \bibinfo{person}{Sarah Kim}, {and} \bibinfo{person}{Kyogu Lee}.} \bibinfo{year}{2022}\natexlab{}.
\newblock \showarticletitle{{Sketching the Expression: Flexible Rendering of Expressive Piano Performance with Self-Supervised Learning}}. In \bibinfo{booktitle}{\emph{Proceedings of the 23rd International Society for Music Information Retrieval Conference {(ISMIR)}}}.
\newblock
\href{https://doi.org/10.5281/zenodo.7342916}{doi:\nolinkurl{10.5281/zenodo.7342916}}


\bibitem[Russell(1980)]%
        {russell1980circumplex}
\bibfield{author}{\bibinfo{person}{James~A Russell}.} \bibinfo{year}{1980}\natexlab{}.
\newblock \showarticletitle{A circumplex model of affect.}
\newblock \bibinfo{journal}{\emph{Journal of personality and social psychology}} \bibinfo{volume}{39}, \bibinfo{number}{6} (\bibinfo{year}{1980}), \bibinfo{pages}{1161}.
\newblock
\href{https://doi.org/10.1037/h0077714}{doi:\nolinkurl{10.1037/h0077714}}


\bibitem[Shazeer(2019)]%
        {shazeer2019fast}
\bibfield{author}{\bibinfo{person}{Noam Shazeer}.} \bibinfo{year}{2019}\natexlab{}.
\newblock \showarticletitle{{Fast Transformer Decoding: One Write-Head is All You Need}}.
\newblock \bibinfo{journal}{\emph{arXiv preprint arXiv:1911.02150}} (\bibinfo{year}{2019}).
\newblock
\href{https://doi.org/10.48550/arXiv.1911.02150}{doi:\nolinkurl{10.48550/arXiv.1911.02150}}


\bibitem[Shazeer(2020)]%
        {shazeer2020glu}
\bibfield{author}{\bibinfo{person}{Noam Shazeer}.} \bibinfo{year}{2020}\natexlab{}.
\newblock \showarticletitle{{GLU Variants Improve Transformer}}.
\newblock \bibinfo{journal}{\emph{arXiv preprint arXiv:2002.05202}} (\bibinfo{year}{2020}).
\newblock
\href{https://doi.org/10.48550/arXiv.2002.05202}{doi:\nolinkurl{10.48550/arXiv.2002.05202}}


\bibitem[Stoica et~al\mbox{.}(2025)]%
        {stoica2025contrastive}
\bibfield{author}{\bibinfo{person}{George Stoica}, \bibinfo{person}{Vivek Ramanujan}, \bibinfo{person}{Xiang Fan}, \bibinfo{person}{Ali Farhadi}, \bibinfo{person}{Ranjay Krishna}, {and} \bibinfo{person}{Judy Hoffman}.} \bibinfo{year}{2025}\natexlab{}.
\newblock \showarticletitle{Contrastive Flow Matching}. In \bibinfo{booktitle}{\emph{IEEE/CVF International Conference on Computer Vision (ICCV)}}.
\newblock


\bibitem[Su et~al\mbox{.}(2024)]%
        {su2024roformer}
\bibfield{author}{\bibinfo{person}{Jianlin Su}, \bibinfo{person}{Murtadha Ahmed}, \bibinfo{person}{Yu Lu}, \bibinfo{person}{Sh~engfeng Pan}, \bibinfo{person}{Wen Bo}, {and} \bibinfo{person}{Yunfeng Liu}.} \bibinfo{year}{2024}\natexlab{}.
\newblock \showarticletitle{{Roformer: Enhanced transformer with rotary position embedding}}.
\newblock \bibinfo{journal}{\emph{Neurocomputing}}  \bibinfo{volume}{568} (\bibinfo{year}{2024}), \bibinfo{pages}{127063}.
\newblock
\href{https://doi.org/10.1016/j.neucom.2023.127063}{doi:\nolinkurl{10.1016/j.neucom.2023.127063}}


\bibitem[Tal et~al\mbox{.}(2024)]%
        {tal2024jasco}
\bibfield{author}{\bibinfo{person}{Or Tal}, \bibinfo{person}{Alon Ziv}, \bibinfo{person}{Itai Gat}, \bibinfo{person}{Felix Kreuk}, {and} \bibinfo{person}{Yossi Adi}.} \bibinfo{year}{2024}\natexlab{}.
\newblock \showarticletitle{Joint audio and symbolic conditioning for temporally controlled text-to-music generation}. In \bibinfo{booktitle}{\emph{Proceedings of the 25th International Society for Music Information Retrieval Conference {(ISMIR)}}}.
\newblock
\href{https://doi.org/10.5281/zenodo.14877325}{doi:\nolinkurl{10.5281/zenodo.14877325}}


\bibitem[Tang et~al\mbox{.}(2025a)]%
        {tang2025towards}
\bibfield{author}{\bibinfo{person}{Jingjing Tang}, \bibinfo{person}{Erica Cooper}, \bibinfo{person}{Xin Wang}, \bibinfo{person}{Junichi Yamagishi}, {and} \bibinfo{person}{György Fazekas}.} \bibinfo{year}{2025}\natexlab{a}.
\newblock \showarticletitle{{Towards An Integrated Approach for Expressive Piano Performance Synthesis from Music Scores}}. In \bibinfo{booktitle}{\emph{Proceeding of the 50th IEEE International Conference on Acoustics, Speech and Signal Processing (ICASSP)}}. \bibinfo{pages}{1--5}.
\newblock
\href{https://doi.org/10.1109/ICASSP49660.2025.10890623}{doi:\nolinkurl{10.1109/ICASSP49660.2025.10890623}}


\bibitem[Tang et~al\mbox{.}(2025b)]%
        {tang2025midivalle}
\bibfield{author}{\bibinfo{person}{Jingjing Tang}, \bibinfo{person}{Xin Wang}, \bibinfo{person}{Zhe Zhang}, \bibinfo{person}{Junichi Yamagishi}, \bibinfo{person}{Geraint Wiggins}, {and} \bibinfo{person}{George Fazekas}.} \bibinfo{year}{2025}\natexlab{b}.
\newblock \showarticletitle{MIDI-VALLE: Improving Expressive Piano Performance Synthesis Through Neural Codec Language Modelling}. In \bibinfo{booktitle}{\emph{Proceedings of the 26th International Society for Music Information Retrieval Conference {(ISMIR)}}}.
\newblock


\bibitem[Tang et~al\mbox{.}(2023)]%
        {tang2023reconstructing}
\bibfield{author}{\bibinfo{person}{Jingjing Tang}, \bibinfo{person}{Geraint Wiggins}, {and} \bibinfo{person}{George Fazekas}.} \bibinfo{year}{2023}\natexlab{}.
\newblock \showarticletitle{Reconstructing Human Expressiveness in Piano Performances with a Transformer Network}. In \bibinfo{booktitle}{\emph{Proceedings of the 16th International Symposium on Computer Music Multidisciplinary Research}}.
\newblock
\href{https://doi.org/10.5281/zenodo.10110378}{doi:\nolinkurl{10.5281/zenodo.10110378}}


\bibitem[Vaswani et~al\mbox{.}(2017)]%
        {vaswani2017attention}
\bibfield{author}{\bibinfo{person}{Ashish Vaswani}, \bibinfo{person}{Noam Shazeer}, \bibinfo{person}{Niki Parmar}, \bibinfo{person}{Jakob Uszkoreit}, \bibinfo{person}{Llion Jones}, \bibinfo{person}{Aidan~N Gomez}, \bibinfo{person}{\L~ukasz Kaiser}, {and} \bibinfo{person}{Illia Polosukhin}.} \bibinfo{year}{2017}\natexlab{}.
\newblock \showarticletitle{{Attention is All you Need}}. In \bibinfo{booktitle}{\emph{Advances in Neural Information Processing Systems {(NIPS)}}}, Vol.~\bibinfo{volume}{30}. \bibinfo{publisher}{Curran Associates, Inc.}, \bibinfo{pages}{5998--6008}.
\newblock


\bibitem[Worrall et~al\mbox{.}(2024)]%
        {worrall2024comparative}
\bibfield{author}{\bibinfo{person}{Kyle Worrall}, \bibinfo{person}{Zongyu Yin}, {and} \bibinfo{person}{Tom Collins}.} \bibinfo{year}{2024}\natexlab{}.
\newblock \showarticletitle{{Comparative Evaluation in the Wild: Systems for the Expressive Rendering of Music}}.
\newblock \bibinfo{journal}{\emph{IEEE Transactions on Artificial Intelligence}} \bibinfo{volume}{5}, \bibinfo{number}{10} (\bibinfo{year}{2024}), \bibinfo{pages}{5290--5303}.
\newblock
\href{https://doi.org/10.1109/TAI.2024.3408717}{doi:\nolinkurl{10.1109/TAI.2024.3408717}}


\bibitem[Xia(2016)]%
        {xia2016expressive}
\bibfield{author}{\bibinfo{person}{Gus~Guangyu Xia}.} \bibinfo{year}{2016}\natexlab{}.
\newblock \emph{\bibinfo{title}{{Expressive Collaborative Music Performance via Machine Learning}}}.
\newblock \bibinfo{thesistype}{Ph.\,D. Dissertation}. \bibinfo{school}{Carnegie Mellon University}.
\newblock


\bibitem[Yan and Duan(2024)]%
        {yan2024transkun}
\bibfield{author}{\bibinfo{person}{Yujia Yan} {and} \bibinfo{person}{Zhiyao Duan}.} \bibinfo{year}{2024}\natexlab{}.
\newblock \showarticletitle{Scoring Time Intervals Using Non-Hierarchical Transformer for Automatic Piano Transcription}. In \bibinfo{booktitle}{\emph{Proceedings of the 25th International Society for Music Information Retrieval Conference {(ISMIR)}}}.
\newblock
\href{https://doi.org/10.5281/zenodo.14877493}{doi:\nolinkurl{10.5281/zenodo.14877493}}


\bibitem[Yang et~al\mbox{.}(2023)]%
        {yang2023diffusion}
\bibfield{author}{\bibinfo{person}{Ling Yang}, \bibinfo{person}{Zhilong Zhang}, \bibinfo{person}{Yang Song}, \bibinfo{person}{Shenda Hong}, \bibinfo{person}{Runsheng Xu}, \bibinfo{person}{Yue Zhao}, \bibinfo{person}{Wentao Zhang}, \bibinfo{person}{Bin Cui}, {and} \bibinfo{person}{Ming-Hsuan Yang}.} \bibinfo{year}{2023}\natexlab{}.
\newblock \showarticletitle{Diffusion models: A comprehensive survey of methods and applications}.
\newblock \bibinfo{journal}{\emph{Comput. Surveys}} \bibinfo{volume}{56}, \bibinfo{number}{4} (\bibinfo{year}{2023}), \bibinfo{pages}{1--39}.
\newblock
\href{https://doi.org/10.1145/3626235}{doi:\nolinkurl{10.1145/3626235}}


\bibitem[Zacharov et~al\mbox{.}(2019)]%
        {zacharov2019zhores}
\bibfield{author}{\bibinfo{person}{Igor Zacharov}, \bibinfo{person}{Rinat Arslanov}, \bibinfo{person}{Maksim Gunin}, \bibinfo{person}{Daniil Stefonishin}, \bibinfo{person}{Andrey Bykov}, \bibinfo{person}{Sergey Pavlov}, \bibinfo{person}{Oleg Panarin}, \bibinfo{person}{Anton Maliutin}, \bibinfo{person}{Sergey Rykovanov}, {and} \bibinfo{person}{Maxim Fedorov}.} \bibinfo{year}{2019}\natexlab{}.
\newblock \showarticletitle{{“Zhores”—Petaflops supercomputer for data-driven modeling, machine learning and artificial intelligence installed in Skolkovo Institute of Science and Technology}}.
\newblock \bibinfo{journal}{\emph{Open Engineering}} \bibinfo{volume}{9}, \bibinfo{number}{1} (\bibinfo{year}{2019}), \bibinfo{pages}{512--520}.
\newblock
\href{https://doi.org/10.1515/eng-2019-0059}{doi:\nolinkurl{10.1515/eng-2019-0059}}


\bibitem[Zeng et~al\mbox{.}(2021)]%
        {zeng2021musicbert}
\bibfield{author}{\bibinfo{person}{Mingliang Zeng}, \bibinfo{person}{Xu Tan}, \bibinfo{person}{Rui Wang}, \bibinfo{person}{Zeqian Ju}, \bibinfo{person}{Tao Qin}, {and} \bibinfo{person}{Tie-Yan Liu}.} \bibinfo{year}{2021}\natexlab{}.
\newblock \showarticletitle{{{MusicBERT}: Symbolic Music Understanding with Large-Scale Pre-Training}}. In \bibinfo{booktitle}{\emph{Findings of the Association for Computational Linguistics: ACL-IJCNLP 2021}}. \bibinfo{pages}{791--800}.
\newblock
\href{https://doi.org/10.18653/v1/2021.findings-acl.70}{doi:\nolinkurl{10.18653/v1/2021.findings-acl.70}}


\bibitem[Zhang et~al\mbox{.}(2024)]%
        {zhang2024dexter}
\bibfield{author}{\bibinfo{person}{Huan Zhang}, \bibinfo{person}{Shreyan Chowdhury}, \bibinfo{person}{Carlos~Eduardo Cancino-Chac{\'o}n}, \bibinfo{person}{Jinhua Liang}, \bibinfo{person}{Simon Dixon}, {and} \bibinfo{person}{Gerhard Widmer}.} \bibinfo{year}{2024}\natexlab{}.
\newblock \showarticletitle{DExter: Learning and Controlling Performance Expression with Diffusion Models}.
\newblock \bibinfo{journal}{\emph{Applied Sciences}} \bibinfo{volume}{14}, \bibinfo{number}{15} (\bibinfo{year}{2024}), \bibinfo{pages}{6543}.
\newblock
\href{https://doi.org/doi.org/10.3390/app14156543}{doi:\nolinkurl{doi.org/10.3390/app14156543}}


\bibitem[Zhang et~al\mbox{.}(2025)]%
        {zhang2025renderbox}
\bibfield{author}{\bibinfo{person}{Huan Zhang}, \bibinfo{person}{Akira Maezawa}, {and} \bibinfo{person}{Simon Dixon}.} \bibinfo{year}{2025}\natexlab{}.
\newblock \showarticletitle{{RenderBox: Expressive Performance Rendering with Text Control}}.
\newblock \bibinfo{journal}{\emph{arXiv preprint arXiv:2502.07711}} (\bibinfo{year}{2025}).
\newblock
\href{https://doi.org/10.48550/arXiv.2502.07711}{doi:\nolinkurl{10.48550/arXiv.2502.07711}}


\bibitem[Zhang et~al\mbox{.}(2022)]%
        {zhang2022atepp}
\bibfield{author}{\bibinfo{person}{Huan Zhang}, \bibinfo{person}{Jingjing Tang}, \bibinfo{person}{Syed Rifat~Mahmud Rafee}, {and} \bibinfo{person}{Simon Dixon~Gy{\"o}rgy Fazekas}.} \bibinfo{year}{2022}\natexlab{}.
\newblock \showarticletitle{{ATEPP: A Dataset of Automatically Transcribed Expressive Piano Performance}}. In \bibinfo{booktitle}{\emph{Proceedings of the 23rd International Society for Music Information Retrieval Conference {(ISMIR)}}}.
\newblock
\href{https://doi.org/10.5281/zenodo.7342764}{doi:\nolinkurl{10.5281/zenodo.7342764}}


\end{thebibliography}

\newpage
\nobalance
\appendix

\setcounter{figure}{0}
\setcounter{table}{0}
\counterwithin{figure}{section}
\counterwithin{table}{section}
\renewcommand\thefigure{\thesection.\arabic{figure}}
\renewcommand\thetable{\thesection.\arabic{table}}
\renewcommand{\thelstlisting}{\thesection.\arabic{lstlisting}}

\section{Limitations}

We acknowledge the limitations of the presented data encoding, model design, and experiments. Regarding data encoding, we do not explicitly model MIDI pedals. The approximation of the sustain effect using a joint pressed and sustain duration prediction allows estimating the pedals based on the prolonged note durations. As discussed in the literature \cite{jeong2019virtuosonet, zhang2024dexter}, complete pedal modeling is challenging, and no existing solutions model them correctly.

Another shortcoming is the way trills are handled. Human performers produce an arbitrary number of notes for trills. During data preprocessing, we match the number of performed notes in trills to the number of notes in the scores, creating a note-by-note match between the scores and performances. In some cases, trills may contain too many notes and sound unnatural. However, MusicXML files in open-source datasets are generally human-edited and may contain false notes and errors, resulting in suboptimal performance. In a side-by-side comparison, though, all models process the same scores, so any errors in the score data are reproduced by each model.

The control capabilities of PianoFlow are limited by the quality of the emotion classifier and the latent text space of the Flan-T5 text encoder. The trained model learned to map the text embeddings to emotions in the music. However, an arbitrary text may produce undesirable results due to its proximity in text space to the discrete emotions used for training. The coherence of the model's responses to emotions and texts should be evaluated in a separate, comprehensive study. Additionally, we did not conduct any ablation studies to determine the impact of different pre-trained text encoders on the quality of emotion-driven training and inference.

All models were trained for 300,000 iterations in the experiments. However, PianoFlow does not overfit to the training dataset and can achieve better performance with additional training steps. The number of training steps was artificially limited to ensure a fair comparison. For instance, the MLM model begins to slowly overfit the training data around 300,000 iterations.

Given the listed limitations, future work should focus on correctly modeling pedal effects, handling trills, and improving the quality and interpretability of text-driven inference.

\section{Emotion Dataset}\label{sec:dataset}

Table~\ref{tab:emotion-labels} provides a list of 33 emotion labels used for the task of classifying emotions in performed piano pieces.

\begin{table}[h!]
% \small
    \caption{A list of emotion labels.}
    \label{tab:emotion-labels}
    \newcolumntype{C}{>{\centering\arraybackslash}X}%
    \newcolumntype{L}{>{\raggedright\arraybackslash}X}%
    \begin{tabularx}{\columnwidth}{LLL}
        \toprule
        anger         & gentle        & mysterious     \\
        anxious       & happy         & nostalgia      \\
        calm          & harsh         & passionately   \\
        capricious    & heavy         & rapidly        \\
        comical       & impetuous     & reflective     \\
        decisive      & important     & religious      \\
        depressed     & kind          & sad            \\
        dreamy        & longing       & sincere        \\
        elegant       & marching      & sleepy         \\
        enthusiastic  & melancholic   & solemn         \\
        fierce        & melodious     & triumphantly   \\
        \bottomrule
    \end{tabularx}
\end{table}

During the training, we sample a template used to describe the emotions in music using text. The templates are used to provide the models with diverse points in the Flan-T5 \cite{chung2024scaling} text embedding space. Listing~\ref{lst:templates} provides a list of 16 used templates.

\begin{lstlisting}[language=Python, caption={Text conditioning templates}, label={lst:templates}]
templates = [
    "{}",
    "{} emotion",
    "{} music",
    "{} music performance",
    "{} music emotion",
    "Musical mood: {}",
    "Play music in a {} mood",
    "Perform music in a {} mood",
    "Play {} music",
    "Perform {} music",
    "Music described as {}",
    "Described as {}",
    "Music classified as {}",
    "Classified as {}",
    "Music performance with a {} emotion",
    "Music performed {}"
]
\end{lstlisting}

\section{Tokenizer and Model Configuration}

Table ~\ref{tab:tokenizer-config} details the configuration of the SyMuPe tokenizer used by the token-based baseline models. It provides a complete list of score and performance tokens, their units of measure and resolution, value range, and number of tokens in the vocabulary.

\begin{table}[ht]
    \centering
    \caption{Configuration of the SyMuPe tokenizer.}
    \label{tab:tokenizer-config}
    % \newcolumntype{C}{>{\centering\arraybackslash}X}%
    % \newcolumntype{L}{>{\raggedright\arraybackslash}X}%
    \resizebox{\columnwidth}{!}{
    \begin{tabular}{ll ccc}
        \toprule
        \textbf{Feature} & \textbf{Resolution} & \textbf{Min} & \textbf{Max} & \textbf{\#Tokens} \\ 
        \midrule
        \multicolumn{5}{c}{Score Tokens} \\
        \midrule
        Pitch & MIDI bins & 21 & 108 & 91 \\
        Position & 96th note & 0 & 192 & 196 \\
        PositionShift & 96th note & 0 & 1536 & 136 \\
        Duration & 96th note & 0 & 1536 & 136 \\
        \midrule
        Tempo & BPM & 15. & 480. & 164 \\
        Velocity & MIDI bins & 0 & 127 & 131 \\
        \midrule
        \multicolumn{5}{c}{Performance Tokens} \\
        \midrule
        Velocity & MIDI bins & 0 & 127 & 131 \\
        TimeShift & seconds & -0.5 & 10. & 365 \\
        TimeDuration & seconds & 0. & 10. & 313 \\
        TimeDurationSustain & seconds & 0. & 10. & 313 \\
        \midrule
        Tempo & BPM & 15. & 480. & 164 \\
        \bottomrule
    \end{tabular}}
\end{table}

Table~\ref{tab:model-config} presents the configurations for the base \texttt{PianoFlow} and the baseline \texttt{MLM} and \texttt{EncDec} models. It lists the transformer configurations, learning objectives, input and output feature types and dimensions, training-related parameters, and inference complexity.

\begin{table*}[ht]
    \centering
    \caption{Detailed configuration of PianoFlow and baseline transformer models.}
    \label{tab:model-config}
    \newcolumntype{C}{>{\centering\arraybackslash}X}%
    \newcolumntype{L}{>{\raggedright\arraybackslash}X}%
    \begin{tabularx}{0.66\textwidth}{lCCC}
        \toprule
        & \textbf{PianoFlow} & \textbf{MLM} & \textbf{EncDec} \\ 
        \midrule
        \multicolumn{4}{c}{Model Configuration} \\
        \midrule
        Transformer Layers & 8 & 8 & 4+4 \\
        Model Dimension & 512 & 512 & 512 \\
        Attention Heads & 8 & 8 & 8 \\
        Feedforward Dimension & 1536 & 1536 & 1536 \\
        Token Embedding Dimension & 64 & 64 & 64 \\
        Time Embedding Dimension & 64 & - & - \\
        Normalization & AdaLayerNorm & LayerNorm & LayerNorm \\
        Learning Objective & OT-CFM & MLM & CLM \\
        \midrule
        \multicolumn{4}{c}{Data Configuration} \\
        \midrule
        Score Features & 4 & 4 & 4\\
        Performance Features & 4 & 4 & 4\\
        Input Feature Type & vectors & tokens & tokens\\
        Score Control Tokens & 2 & 2 & 2\\
        Performance Control Tokens & 1 & 1 & 1\\
        Text Embedding Dimension & 768 & 768 & 768\\
        Classifier-Free Guidance Dropout & 0.2 & 0.2 & 0.2 \\
        \midrule
        \multicolumn{4}{c}{Training Configuration} \\
        \midrule
        Maximum Sequence Length & 256 notes & 256 notes & 256 notes\\
        Batch Size & 128 & 128 & 128 \\
        Mask Ratios & $\mathcal{U}[0.1, 0.9]$ & $\mathcal{U}[0.1, 0.9]$ & $\mathcal{U}[0.1, 0.9]$ \\
        Initial Learning Rate & $2 \cdot 10^{-4}$ & $2 \cdot 10^{-4}$ & $2 \cdot 10^{-4}$ \\
        Final Learning Rate & $10^{-4}$ & $10^{-4}$ & $10^{-4}$ \\ 
        Warmup Steps & 1000 & 1000 & 1000\\
        \midrule
        \multicolumn{4}{c}{Inference Configuration} \\
        \midrule
        Inference Type & iterative & single-step & autoregressive \\
        Complexity & $\mathcal{O}(k)$ & $\mathcal{O}(1)$ & $\mathcal{O}(n)$ \\
        \bottomrule
    \end{tabularx}
\end{table*}

\section{Ablation Study}

Table~\ref{tab:ablation-evaluation} provides a complete objective evaluation of all model configurations and changes examined during the experiments. While the objective metrics do not fully translate into the perception of the audio samples, they still provide an indication of how each configuration might perform. The final configuration described in the paper was chosen based on internal perception, as evaluating all model combinations using a survey is time consuming.

\begin{table*}[ht]
    \caption{A complete statistical comparison of the rendered and human performances as part of the ablation study. \texttt{Vel} - velocity, \texttt{IOI} - inter-onset-interval, \texttt{OD} - relative onset deviation, \texttt{Art} - articulation, \texttt{ArtS} - sustained articulation. \texttt{Param} - number of parameters, \texttt{NPS} - average number of notes rendered per second on V100 GPU.}
    \label{tab:ablation-evaluation}
    \newcolumntype{C}{>{\centering\arraybackslash}X}%
    \newcolumntype{L}{>{\raggedright\arraybackslash}X}%
    \resizebox{\textwidth}{!}{\begin{tabularx}{\textwidth}{lCCcCCCCCcCCCCC}
        \toprule
        &&&& \multicolumn{5}{c}{\textbf{Pearson’s Correlation ($\uparrow$)}} && \multicolumn{5}{c}{\textbf{KL Divergence ($\downarrow$)}} \\
        \cmidrule{5-9} \cmidrule{11-15}
        \textbf{Model} & \textbf{Param} & \textbf{NPS} && \textbf{Vel} & \textbf{IOI} & \textbf{OD} & \textbf{Art} & \textbf{ArtS} && \textbf{Vel} & \textbf{IOI} & \textbf{OD} & \textbf{Art} & \textbf{ArtS} \\
        \midrule
        Dataset & - & - && 0.65{\scriptsize $\pm$0.18} & 0.90{\scriptsize $\pm$0.11} & 0.24{\scriptsize $\pm$0.18} & 0.50{\scriptsize $\pm$0.14} & 0.46{\scriptsize $\pm$0.20} & & 0.27{\scriptsize $\pm$0.34} & 0.20{\scriptsize $\pm$0.25} & 0.13{\scriptsize $\pm$0.12} & 0.12{\scriptsize $\pm$0.10} & 0.15{\scriptsize $\pm$0.13} \\
        \midrule
        VirtuosoNet \cite{jeong2019graph} & 5M & 1700 && 0.48{\scriptsize $\pm$0.27} & 0.58{\scriptsize $\pm$0.29} & 0.00{\scriptsize $\pm$0.05} & 0.27{\scriptsize $\pm$0.17} & - & & 0.73{\scriptsize $\pm$0.76} & 0.93{\scriptsize $\pm$1.08} & 1.57{\scriptsize $\pm$1.68} & 1.53{\scriptsize $\pm$1.47} &  - \\
        DExter \cite{zhang2024dexter} & 62M & 30 && 0.24{\scriptsize $\pm$0.30} & 0.36{\scriptsize $\pm$0.35} & 0.01{\scriptsize $\pm$0.04} & 0.14{\scriptsize $\pm$0.19} & - & & 1.67{\scriptsize $\pm$1.11} & 0.47{\scriptsize $\pm$0.38} & 0.45{\scriptsize $\pm$0.33} & 0.81{\scriptsize $\pm$0.52} &  - \\
        \midrule
        MLM & 24M & 3450 && 0.41{\scriptsize $\pm$0.18} & 0.80{\scriptsize $\pm$0.13} & 0.07{\scriptsize $\pm$0.07} & 0.30{\scriptsize $\pm$0.10} & 0.30{\scriptsize $\pm$0.14} & & 0.38{\scriptsize $\pm$0.45} & 0.31{\scriptsize $\pm$0.22} & 0.19{\scriptsize $\pm$0.10} & 0.23{\scriptsize $\pm$0.30} & 0.30{\scriptsize $\pm$0.37} \\
        \quad w/o $c_s$ & 24M & 3450 && 0.35{\scriptsize $\pm$0.17} & 0.79{\scriptsize $\pm$0.13} & 0.07{\scriptsize $\pm$0.06} & 0.28{\scriptsize $\pm$0.10} & 0.29{\scriptsize $\pm$0.13} & & 0.37{\scriptsize $\pm$0.40} & 0.39{\scriptsize $\pm$0.32} & 0.20{\scriptsize $\pm$0.11} & 0.28{\scriptsize $\pm$0.35} & 0.37{\scriptsize $\pm$0.43} \\
        \quad w/o SinVE & 24M & 3450 && 0.42{\scriptsize $\pm$0.17} & 0.81{\scriptsize $\pm$0.14} & 0.08{\scriptsize $\pm$0.08} & 0.31{\scriptsize $\pm$0.10} & 0.31{\scriptsize $\pm$0.14} & & 0.37{\scriptsize $\pm$0.41} & 0.37{\scriptsize $\pm$0.59} & 0.21{\scriptsize $\pm$0.18} & 0.26{\scriptsize $\pm$0.34} & 0.31{\scriptsize $\pm$0.56} \\
        \midrule
        EncDec & 24M & 80 && 0.37{\scriptsize $\pm$0.15} & 0.80{\scriptsize $\pm$0.13} & 0.07{\scriptsize $\pm$0.09} & 0.31{\scriptsize $\pm$0.11} & 0.32{\scriptsize $\pm$0.14} & & 0.57{\scriptsize $\pm$0.63} & 0.67{\scriptsize $\pm$0.86} & 0.16{\scriptsize $\pm$0.11} & 0.47{\scriptsize $\pm$0.61} & 0.35{\scriptsize $\pm$0.44} \\
        \quad w/o & 24M & 80 && 0.35{\scriptsize $\pm$0.16} & 0.80{\scriptsize $\pm$0.13} & 0.06{\scriptsize $\pm$0.09} & 0.30{\scriptsize $\pm$0.11} & 0.32{\scriptsize $\pm$0.14} & & 0.59{\scriptsize $\pm$0.60} & 0.67{\scriptsize $\pm$0.79} & 0.15{\scriptsize $\pm$0.11} & 0.48{\scriptsize $\pm$0.61} & 0.37{\scriptsize $\pm$0.48} \\
        \quad w/o SinE & 24M & 80 && 0.35{\scriptsize $\pm$0.14} & 0.79{\scriptsize $\pm$0.14} & 0.05{\scriptsize $\pm$0.05} & 0.30{\scriptsize $\pm$0.10} & 0.30{\scriptsize $\pm$0.13} & & 0.69{\scriptsize $\pm$0.66} & 0.59{\scriptsize $\pm$0.67} & 0.19{\scriptsize $\pm$0.16} & 0.38{\scriptsize $\pm$0.48} & 0.31{\scriptsize $\pm$0.40} \\
        \midrule
        \textbf{PianoFlow} & 24M & 355 && 0.43{\scriptsize $\pm$0.17} & 0.86{\scriptsize $\pm$0.12} & 0.09{\scriptsize $\pm$0.12} & 0.35{\scriptsize $\pm$0.12} & 0.35{\scriptsize $\pm$0.17} & & 0.64{\scriptsize $\pm$0.56} & 0.56{\scriptsize $\pm$0.76} & 0.55{\scriptsize $\pm$0.51} & 0.46{\scriptsize $\pm$0.50} & 0.48{\scriptsize $\pm$0.89} \\
        \quad w/o $c_s$ & 24M & 355 && 0.35{\scriptsize $\pm$0.16} & 0.84{\scriptsize $\pm$0.13} & 0.09{\scriptsize $\pm$0.13} & 0.32{\scriptsize $\pm$0.11} & 0.34{\scriptsize $\pm$0.17} & & 0.74{\scriptsize $\pm$0.59} & 0.99{\scriptsize $\pm$1.14} & 0.56{\scriptsize $\pm$0.53} & 0.75{\scriptsize $\pm$0.70} & 0.75{\scriptsize $\pm$1.14} \\
        \quad w/o SinE & 24M & 355 && 0.42{\scriptsize $\pm$0.17} & 0.86{\scriptsize $\pm$0.12} & 0.09{\scriptsize $\pm$0.13} & 0.34{\scriptsize $\pm$0.11} & 0.33{\scriptsize $\pm$0.17} & & 0.76{\scriptsize $\pm$0.63} & 0.63{\scriptsize $\pm$1.28} & 0.59{\scriptsize $\pm$0.51} & 0.52{\scriptsize $\pm$0.56} & 0.44{\scriptsize $\pm$0.69} \\
        \quad w/o IntEmb & 24M & 355 && 0.45{\scriptsize $\pm$0.17} & 0.85{\scriptsize $\pm$0.13} & 0.08{\scriptsize $\pm$0.10} & 0.34{\scriptsize $\pm$0.12} & 0.35{\scriptsize $\pm$0.18} & & 0.67{\scriptsize $\pm$0.63} & 0.50{\scriptsize $\pm$1.00} & 0.42{\scriptsize $\pm$0.40} & 0.37{\scriptsize $\pm$0.48} & 0.40{\scriptsize $\pm$0.92} \\
        \quad CleanNoteLoss & 24M & 355 && 0.43{\scriptsize $\pm$0.16} & 0.85{\scriptsize $\pm$0.13} & 0.09{\scriptsize $\pm$0.12} & 0.34{\scriptsize $\pm$0.11} & 0.34{\scriptsize $\pm$0.18} & & 0.74{\scriptsize $\pm$0.55} & 0.59{\scriptsize $\pm$1.14} & 0.59{\scriptsize $\pm$0.52} & 0.44{\scriptsize $\pm$0.54} & 0.45{\scriptsize $\pm$0.91} \\
        \midrule
        \multicolumn{10}{l}{\textbf{PianoFlow}, model size} \\
        \quad dim=786, h=12 & 84M & 265 && 0.36{\scriptsize $\pm$0.18} & 0.85{\scriptsize $\pm$0.14} & 0.08{\scriptsize $\pm$0.11} & 0.33{\scriptsize $\pm$0.12} & 0.27{\scriptsize $\pm$0.16} & & 0.90{\scriptsize $\pm$0.64} & 0.70{\scriptsize $\pm$1.56} & 1.05{\scriptsize $\pm$0.64} & 0.61{\scriptsize $\pm$0.60} & 0.56{\scriptsize $\pm$0.85}  \\
        \quad dim=384, h=6 & 10M & 430 && 0.40{\scriptsize $\pm$0.18} & 0.83{\scriptsize $\pm$0.14} & 0.07{\scriptsize $\pm$0.12} & 0.32{\scriptsize $\pm$0.12} & 0.33{\scriptsize $\pm$0.18} & & 0.78{\scriptsize $\pm$0.85} & 0.59{\scriptsize $\pm$0.70} & 0.56{\scriptsize $\pm$0.43} & 0.53{\scriptsize $\pm$0.89} & 0.39{\scriptsize $\pm$0.95} \\
        \quad dim=256, h=4 & 3M & 560 && 0.39{\scriptsize $\pm$0.18} & 0.85{\scriptsize $\pm$0.12} & 0.07{\scriptsize $\pm$0.08} & 0.32{\scriptsize $\pm$0.11} & 0.35{\scriptsize $\pm$0.18} & & 0.80{\scriptsize $\pm$0.73} & 0.52{\scriptsize $\pm$0.72} & 0.51{\scriptsize $\pm$0.48} & 0.64{\scriptsize $\pm$0.73} & 0.52{\scriptsize $\pm$0.86} \\
        \midrule
        \multicolumn{10}{l}{\textbf{PianoFlow}, inference steps $k$ and adaptive step factor $\gamma$} \\
        \quad $k=1$ & 24M & 3200 && 0.51{\scriptsize $\pm$0.18} & 0.84{\scriptsize $\pm$0.15} & 0.11{\scriptsize $\pm$0.15} & 0.41{\scriptsize $\pm$0.15} & 0.44{\scriptsize $\pm$0.16} & & 0.99{\scriptsize $\pm$1.09} & 0.74{\scriptsize $\pm$1.01} & 0.45{\scriptsize $\pm$0.51} & 1.20{\scriptsize $\pm$1.28} & 1.20{\scriptsize $\pm$1.38} \\
        \quad $k=4$ & 24M & 885 && 0.45{\scriptsize $\pm$0.17} & 0.86{\scriptsize $\pm$0.12} & 0.10{\scriptsize $\pm$0.15} & 0.37{\scriptsize $\pm$0.12} & 0.37{\scriptsize $\pm$0.17} & & 0.87{\scriptsize $\pm$0.79} & 0.66{\scriptsize $\pm$1.01} & 1.11{\scriptsize $\pm$1.02} & 0.74{\scriptsize $\pm$0.82} & 0.71{\scriptsize $\pm$1.30} \\
        \quad $k=10, \gamma = 0.75$ & 24M & 355 && 0.43{\scriptsize $\pm$0.17} & 0.86{\scriptsize $\pm$0.12} & 0.09{\scriptsize $\pm$0.12} & 0.35{\scriptsize $\pm$0.12} & 0.35{\scriptsize $\pm$0.17} & & 0.64{\scriptsize $\pm$0.56} & 0.56{\scriptsize $\pm$0.76} & 0.55{\scriptsize $\pm$0.51} & 0.46{\scriptsize $\pm$0.50} & 0.48{\scriptsize $\pm$0.89} \\
        \quad $k=10, \gamma = 1.$ & 24M & 355 && 0.41{\scriptsize $\pm$0.16} & 0.85{\scriptsize $\pm$0.12} & 0.10{\scriptsize $\pm$0.14} & 0.33{\scriptsize $\pm$0.11} & 0.32{\scriptsize $\pm$0.17} & & 0.66{\scriptsize $\pm$0.59} & 0.64{\scriptsize $\pm$0.93} & 0.84{\scriptsize $\pm$0.80} & 0.51{\scriptsize $\pm$0.46} & 0.48{\scriptsize $\pm$0.78} \\
        \quad $k=20, \gamma = 0.75$ & 24M & 180 && 0.43{\scriptsize $\pm$0.17} & 0.86{\scriptsize $\pm$0.12} & 0.09{\scriptsize $\pm$0.12} & 0.35{\scriptsize $\pm$0.12} & 0.36{\scriptsize $\pm$0.17} & & 0.58{\scriptsize $\pm$0.52} & 0.51{\scriptsize $\pm$0.62} & 0.32{\scriptsize $\pm$0.25} & 0.41{\scriptsize $\pm$0.46} & 0.45{\scriptsize $\pm$0.83} \\
        \quad $k=20, \gamma = 1.$ & 24M & 180 && 0.40{\scriptsize $\pm$0.16} & 0.84{\scriptsize $\pm$0.13} & 0.09{\scriptsize $\pm$0.12} & 0.32{\scriptsize $\pm$0.11} & 0.31{\scriptsize $\pm$0.16} & & 0.54{\scriptsize $\pm$0.48} & 0.62{\scriptsize $\pm$0.87} & 0.59{\scriptsize $\pm$0.61} & 0.40{\scriptsize $\pm$0.37} & 0.42{\scriptsize $\pm$0.66} \\
        \midrule
        \multicolumn{10}{l}{\textbf{PianoFlow}, training dataset} \\
        \quad (n)ASAP \cite{peter2023nasap} & 24M & 355 && 0.36{\scriptsize $\pm$0.18} & 0.84{\scriptsize $\pm$0.12} & 0.09{\scriptsize $\pm$0.10} & 0.32{\scriptsize $\pm$0.12} & 0.31{\scriptsize $\pm$0.15} & & 0.33{\scriptsize $\pm$0.36} & 0.75{\scriptsize $\pm$1.47} & 0.25{\scriptsize $\pm$0.21} & 0.73{\scriptsize $\pm$1.16} & 0.57{\scriptsize $\pm$1.14} \\
        \quad ATEPP \cite{zhang2022atepp} & 24M & 355 && 0.42{\scriptsize $\pm$0.19} & 0.85{\scriptsize $\pm$0.12} & 0.09{\scriptsize $\pm$0.10} & 0.32{\scriptsize $\pm$0.11} & 0.27{\scriptsize $\pm$0.12} & & 0.62{\scriptsize $\pm$0.63} & 0.83{\scriptsize $\pm$1.66} & 0.31{\scriptsize $\pm$0.28} & 0.56{\scriptsize $\pm$0.78} & 0.56{\scriptsize $\pm$0.83} \\
        \quad PERiScoPe-os & 24M & 355 && 0.43{\scriptsize $\pm$0.19} & 0.85{\scriptsize $\pm$0.11} & 0.09{\scriptsize $\pm$0.13} & 0.36{\scriptsize $\pm$0.12} & 0.34{\scriptsize $\pm$0.14} & & 0.58{\scriptsize $\pm$0.60} & 0.61{\scriptsize $\pm$1.41} & 0.21{\scriptsize $\pm$0.14} & 0.50{\scriptsize $\pm$0.65} & 0.38{\scriptsize $\pm$0.68} \\
        \bottomrule
    \end{tabularx}}
\end{table*}

Taking some notes from the ablation evaluation, score control tokens $c_s$ with score tempo and dynamics improve the correlation between the rendered and real performances. Without the sinusoidal embeddings for all token types (\texttt{SinVE}), the models show similar performance within a percentage margin. Additionally, in the models we label all notes interpolated during data preprocessing with a learned binary embedding (\texttt{IntEmb}). While the objective metrics do not capture the difference, it is quite noticeable to the ear. By explicitly conditioning the model with the data on which notes were artificial (and sometimes deadpan), the model produces less deadpan and more expressive music. The result is also better than masking all unplayed notes and computing a loss of only the original clean notes (\texttt{CleanNoteLoss}).

The model size and architecture were chosen to meet the requirements of near-real-time inference on the CPU and real-time inference on low-end GPUs. For both the larger and smaller architectures, the validation loss is worse after 300,000 training steps. However, in the case of the larger model, it outperforms the base model in a few thousand additional training iterations. For inference, the adaptive decreasing step size improves the quality of inference. Interestingly, a model with a single flow-matching inference step shows the highest correlation with the dataset performances. The performances are closer to the mean values of the features in the data, but are less expressive and pleasant to listen to in action. This adds to the need to develop and establish more robust objective metrics for evaluating expressive music performances.

In terms of inference speed, PianoFlow can render up to 355 notes per second on a V100 GPU, which is about 20 to 40 seconds of performed music. So the model works faster than real time on the GPU. On the CPU, the inference is close to real time, and exact real time inference can be achieved with fewer sampling steps $k$, balancing quality and speed.

\begin{figure*}[h!]
	\centering
	\includegraphics[alt={},width=0.76\textwidth]{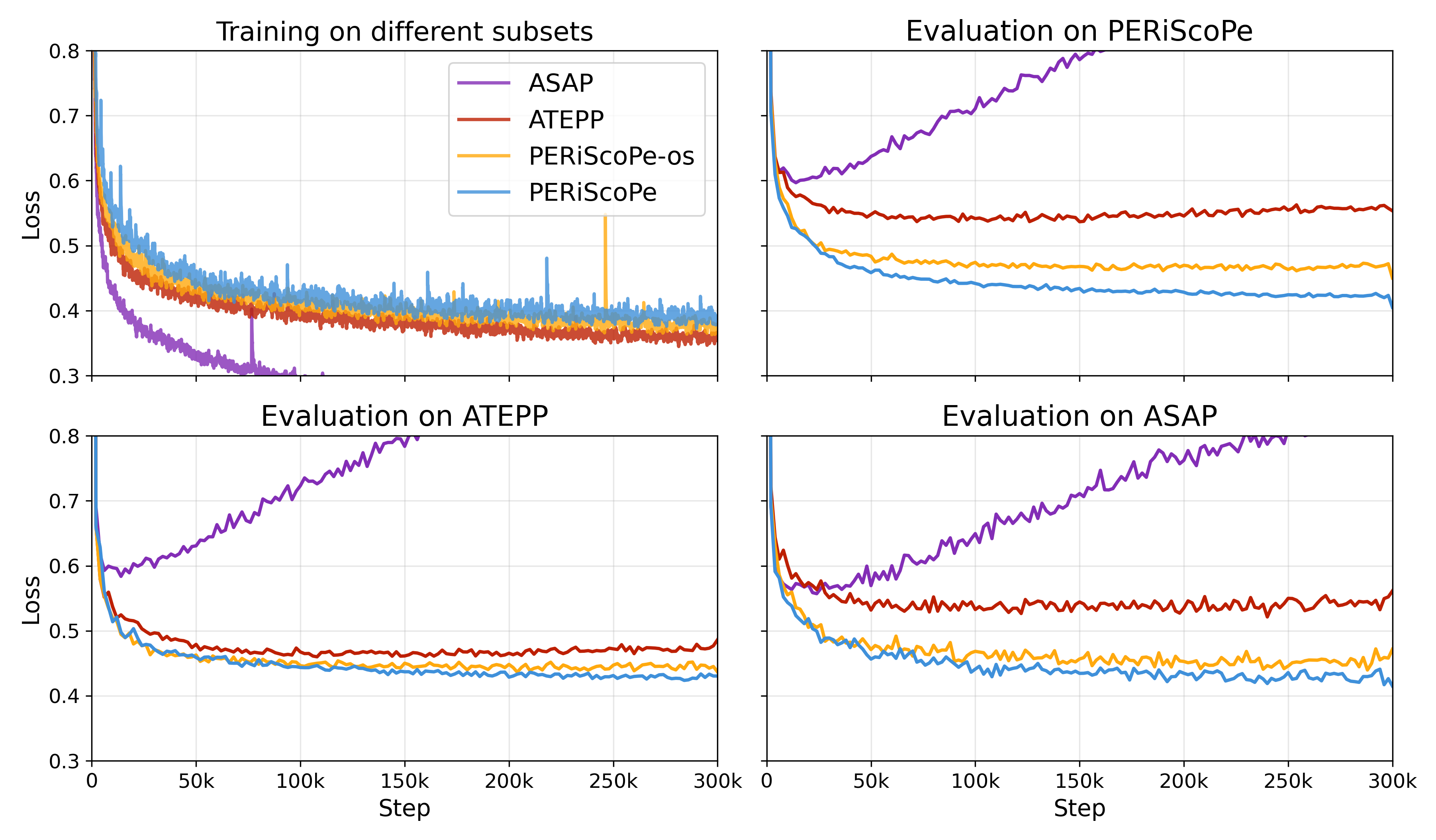}
	\caption{Loss curves for \mbox{PianoFlow} trained on 4 different training subsets (top left) and evaluated on 3 evaluation sets. PERiScoPe-os combines all performances in ASAP \cite{peter2023nasap} and ATEPP \cite{zhang2022atepp} with scores from ASAP, ATEPP and PDMX \cite{long2025pdmx}.}
	\label{fig:training-curves}
\end{figure*}

Figure~\ref{fig:training-curves} shows the training convergence of PianoFlow on different subsets of training data. With smaller, less diverse datasets, PianoFlow begins to overfit the training data. Even with a 24-million-parameter model, overfitting the existing largest open-source ASAP and ATEPP datasets is not that hard. The objective evaluation also shows that models trained on more diverse datasets generalize better. This demonstrates the importance of large-scale, aligned datasets for expressive performance rendering task. This is something that is often overlooked in the literature.

\section{Listening Test}

Table~\ref{tab:eval-pieces} lists all the composers and compositions used in the side-by-side blind listening test. The compositions were chosen to cover different historical musical periods, composers, and musical contexts, ranging from dynamic and expressive to minimalist and quiet. Additionally, we highlight the sources of the human performances used as references. As mentioned in the main paper, only the performances in the ASAP dataset were recorded directly as MIDI, while all others were transcribed from audio to MIDI.

\begin{table*}[ht]
    \caption{A list of musical pieces and sources of human MIDI performances used for blind listening evaluation.}
    \label{tab:eval-pieces}
    % \resizebox{\textwidth}{!}{
    \begin{tabular}{cllc}
        \toprule
        \textbf{Index} & \textbf{Composer} & \textbf{Musical Piece} & \textbf{Performances} \\
        \midrule
        1 & Johann Sebastian Bach & Fantasia and Fugue in A Minor, BWV 944 & ATEPP \\
        2 & Johann Sebastian Bach & Invention No.11 in G minor, BWV782 & ATEPP \\
        3 & George Frideric Handel & Suite No.7 in G Minor, VI. Passacaille & Proprietary \\
        4 & Ludwigvan Beethoven & Piano Sonata No.25 in G, Op.79 (Andante) & ATEPP \\
        5 & Ludwig van Beethoven & Piano Sonata No.30 in E Major, Op.109 (3rd Movement) & ATEPP \\
        6 & Wolfgang Amadeus Mozart & Piano Sonata No.11 in A Major, K.331 (3rd Movement: Alla Turca) & ASAP/ATEPP\\
        7 & Franz Schubert & Valses sentimentales, D.779 (Waltz in A major) & Proprietary \\
        8 & Frederic Chopin & Ballade No.1 in G Minor, Op.23 & ASAP \\
        9 & Frederic Chopin & Nocturne No.1 in B‑flat Minor, Op.9 No.1 & ATEPP \\
        10 & Franz Liszt & Ballade No.2, S.171 & ASAP \\
        11 & Franz Liszt & Hungarian Rhapsody No.6 & ASAP \\
        12 & Mikhail Glinka & ``The Lark'' from A Farewell to Saint Petersburg & ASAP \\
        13 & Maurice Ravel & Jeux d'eau, M.30 & ATEPP \\
        14 & Erik Satie & 3 Gymnopédies (I. Lent et douloureux) & Proprietary \\
        15 & Sergei Prokofiev & Toccata, Op.11 & ASAP \\
        16 & Edvard Grieg & Piano Sonata, Op.7 (Andante molto) & ATEPP \\
        17 & Edward Elgar & Salut d'amour, Op.12 & ATEPP \\
        18 & Scott Joplin & Maple Leaf Rag & Proprietary \\
        19 & Mily Balakirev & Islamey, Op.18 & ASAP \\
        20 & Sergei Rachmaninoff & Piano Sonata No.2 in B‑flat Minor, Op.36 (Allegro molto) & ATEPP \\
        \bottomrule
    \end{tabular}
    % }
\end{table*}

\begin{figure*}[ht]
	\centering
	\includegraphics[alt={},width=1.\textwidth]{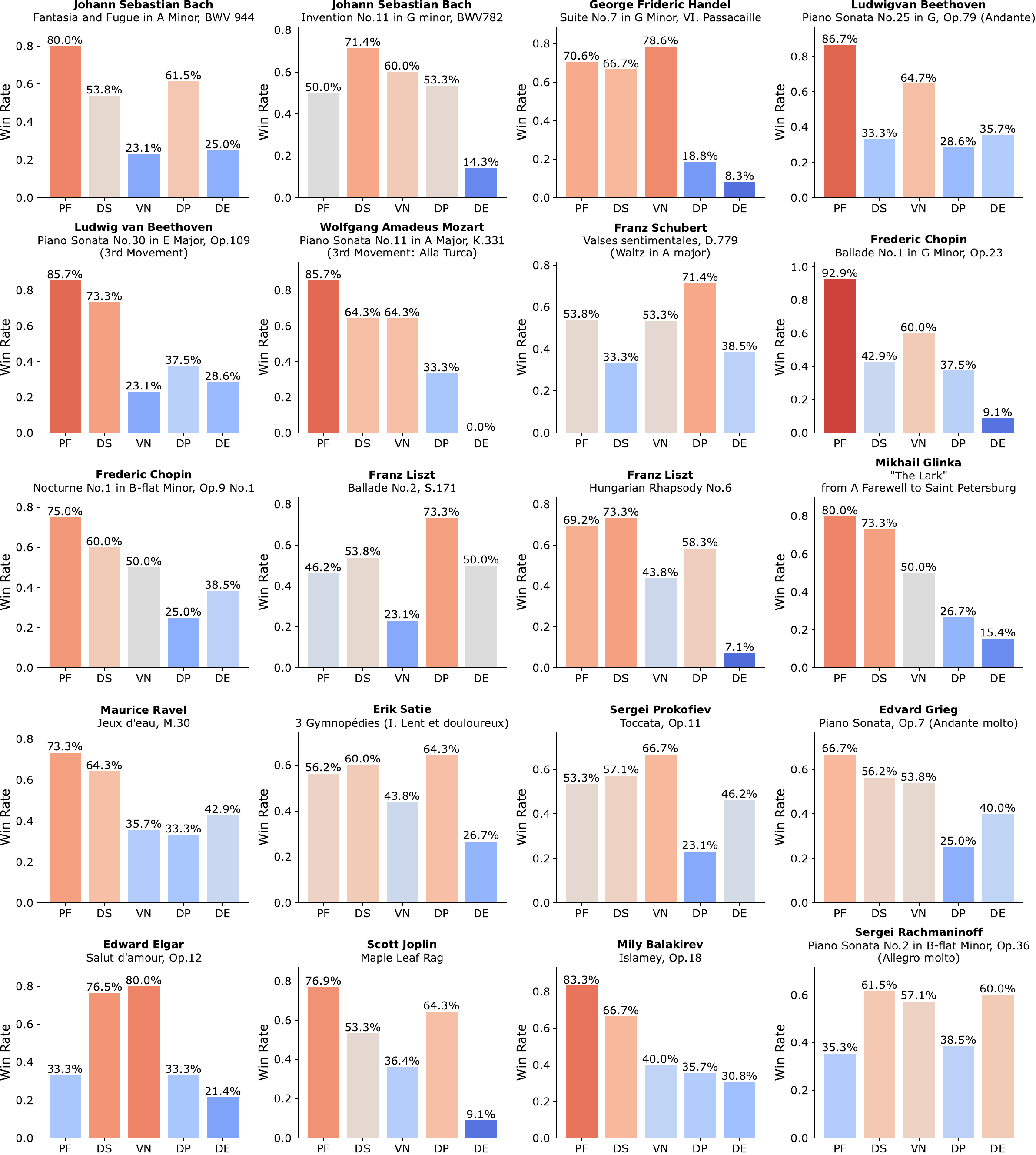}
	\caption{Distribution of votes for each musical piece from the side-by-side listening evaluation. \texttt{PF} - PianoFlow, \texttt{DS} - Dataset, \texttt{VN} - VirtuosoNet, \texttt{DP} - Deadpan, \texttt{DE} - DExter.}
	\label{fig:survey-compositions}
\end{figure*}

Figure~\ref{fig:survey-compositions} shows the listening test results broken down by individual musical composition. Although these statistics are noisier than the aggregated model scores, some observations can still be made. For example, no model excels in all musical contexts. \texttt{PianoFlow} was preferred more for eleven pieces, while \texttt{Dataset}, \texttt{VirtuosoNet} and \texttt{Deadpan} performances were preferred for three distinct compositions each.

A preference for deadpan was found in the contrasting \textit{``Ballade No.2''} by Franz Liszt, a minimalist \textit{``Gymnopedie''} by Erik Satie, and the introduction to \textit{``Waltz in A Major''} by Franz Schubert. These findings suggest that performances, whether real or rendered, are not always perceived as more pleasant than raw score renderings.

Large differences in scores between PianoFlow and VirtuosoNet occurred when the models produced very different tempos. Examples include \textit{``Piano Sonata No. 30''} by Ludwig van Beethoven, \textit{``Fantasia and Fugue in A Minor''} by Johann Sebastian Bach, and \textit{``Maple Leaf Rag''} by Scott Joplin). For Frederic Chopin's \textit{``Ballade No. 1 in G Minor''}, which received the highest PianoFlow scores, the selections seemed to be driven by its slower performance tempo, although the score indicates a faster tempo typically played by professional performers.

\begin{figure*}[ht]
    \centering
    \begin{subfigure}[b]{0.24\textwidth}
        \includegraphics[width=\textwidth]{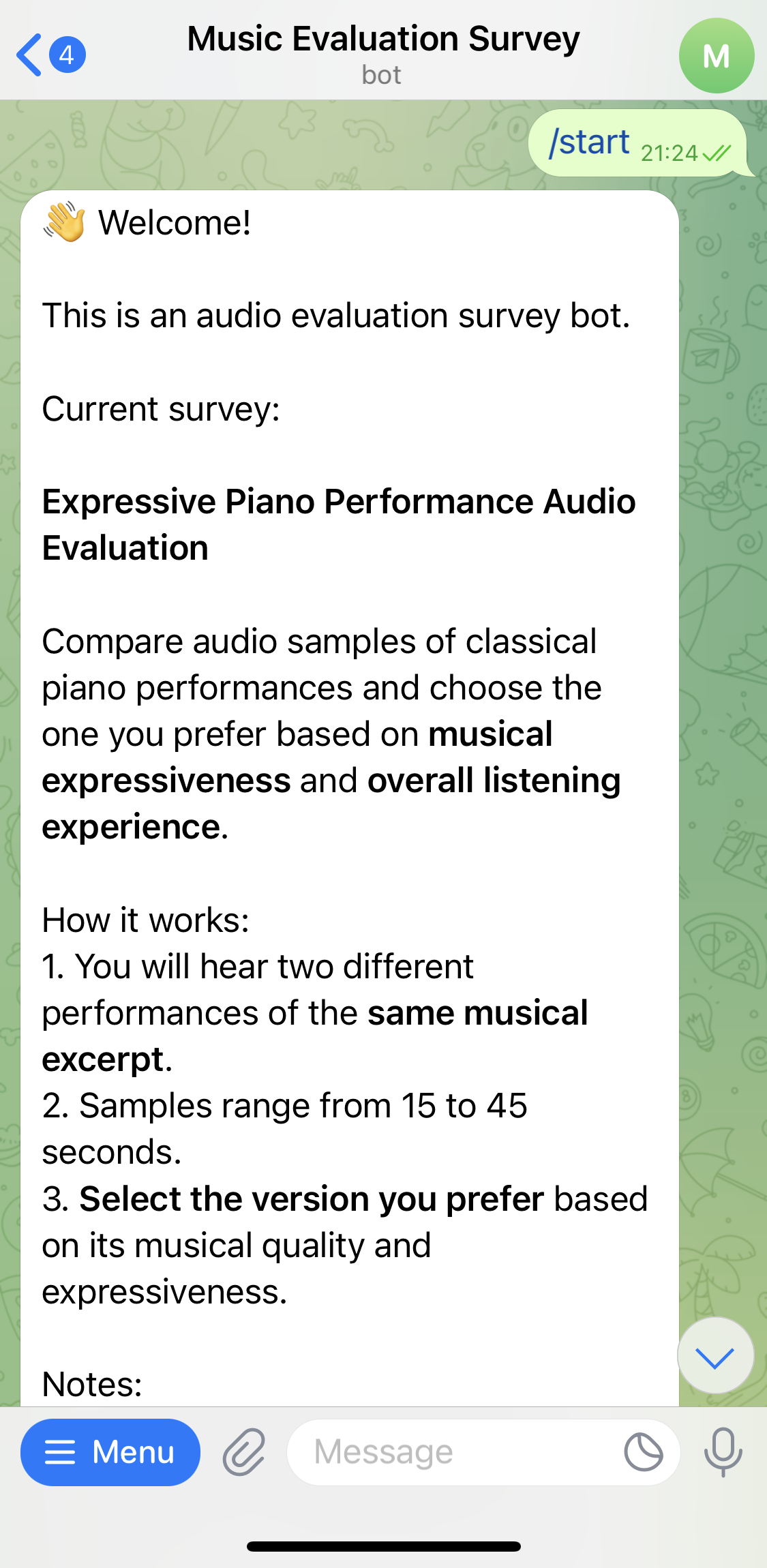}
        \caption{Welcome message, pt.1\\\quad}
        \label{fig:sub1}
    \end{subfigure}
    \hfill
    \begin{subfigure}[b]{0.24\textwidth}
        \includegraphics[width=\textwidth]{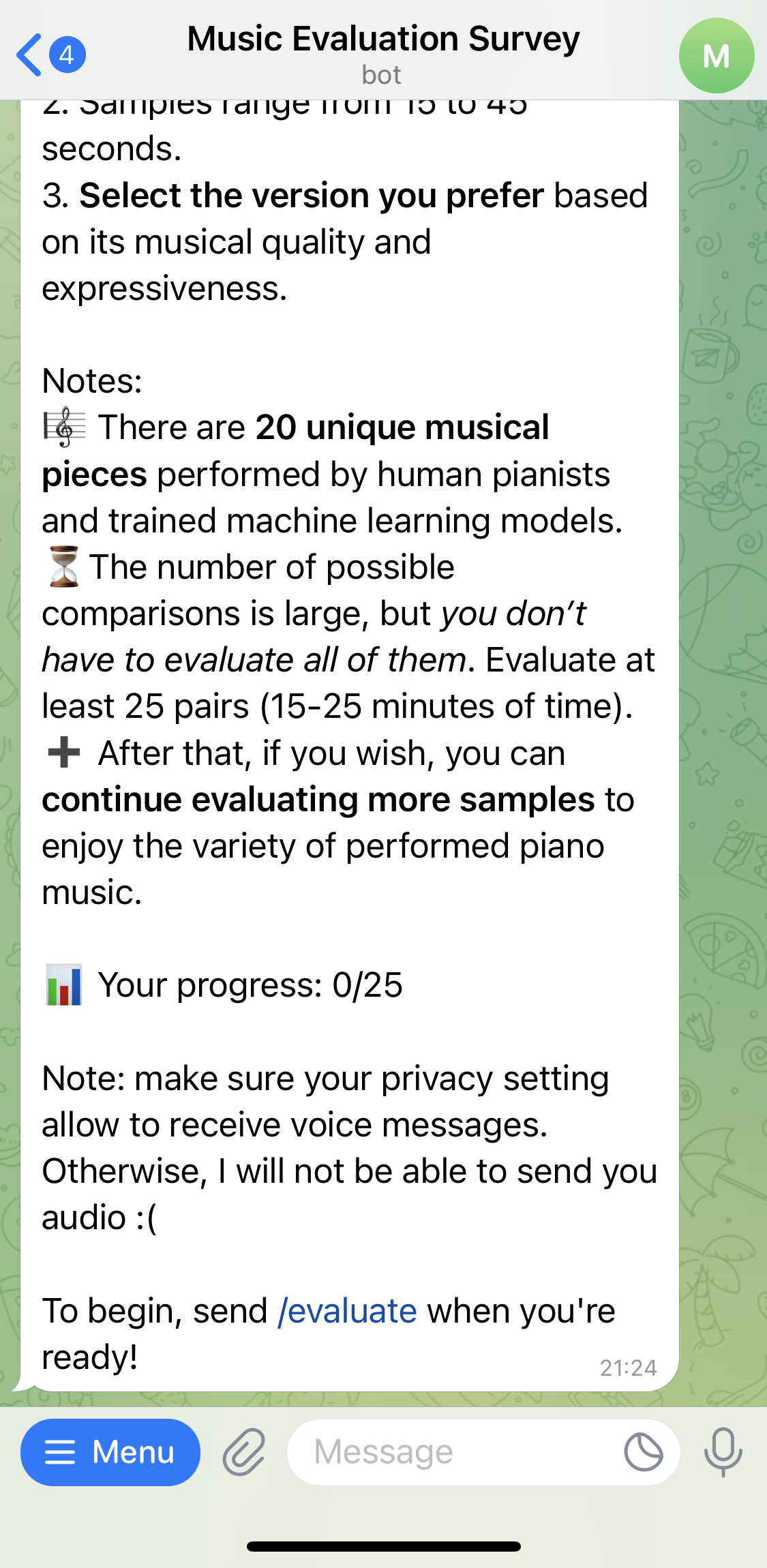}
        \caption{Welcome message, pt.2\\\quad}
        \label{fig:sub2}
    \end{subfigure}
    \hfill
    \begin{subfigure}[b]{0.24\textwidth}
        \includegraphics[width=\textwidth]{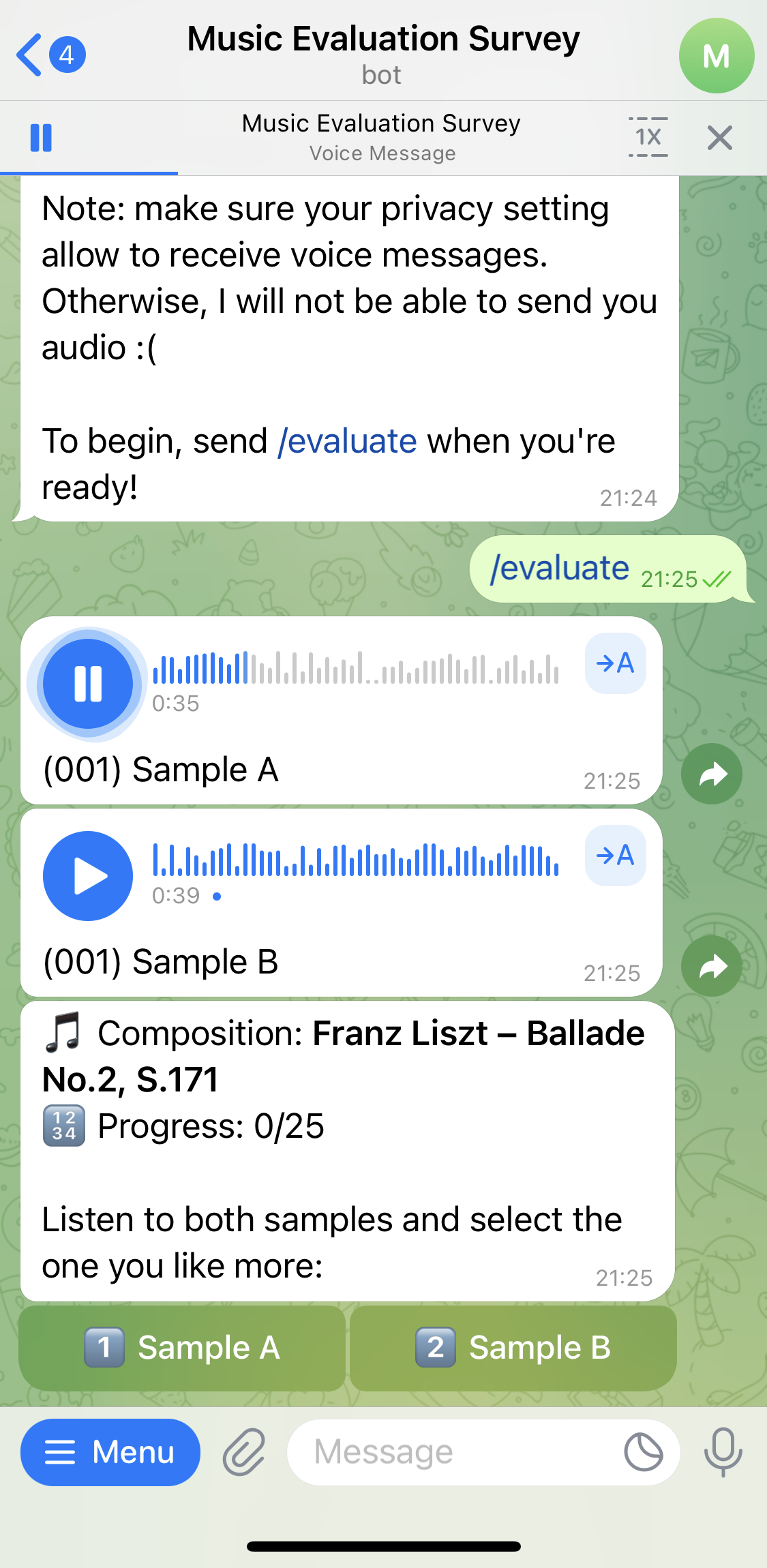}
        \caption{Evaluation process for a pair of samples}
        \label{fig:sub3}
    \end{subfigure}
    \hfill
    \begin{subfigure}[b]{0.24\textwidth}
        \includegraphics[width=\textwidth]{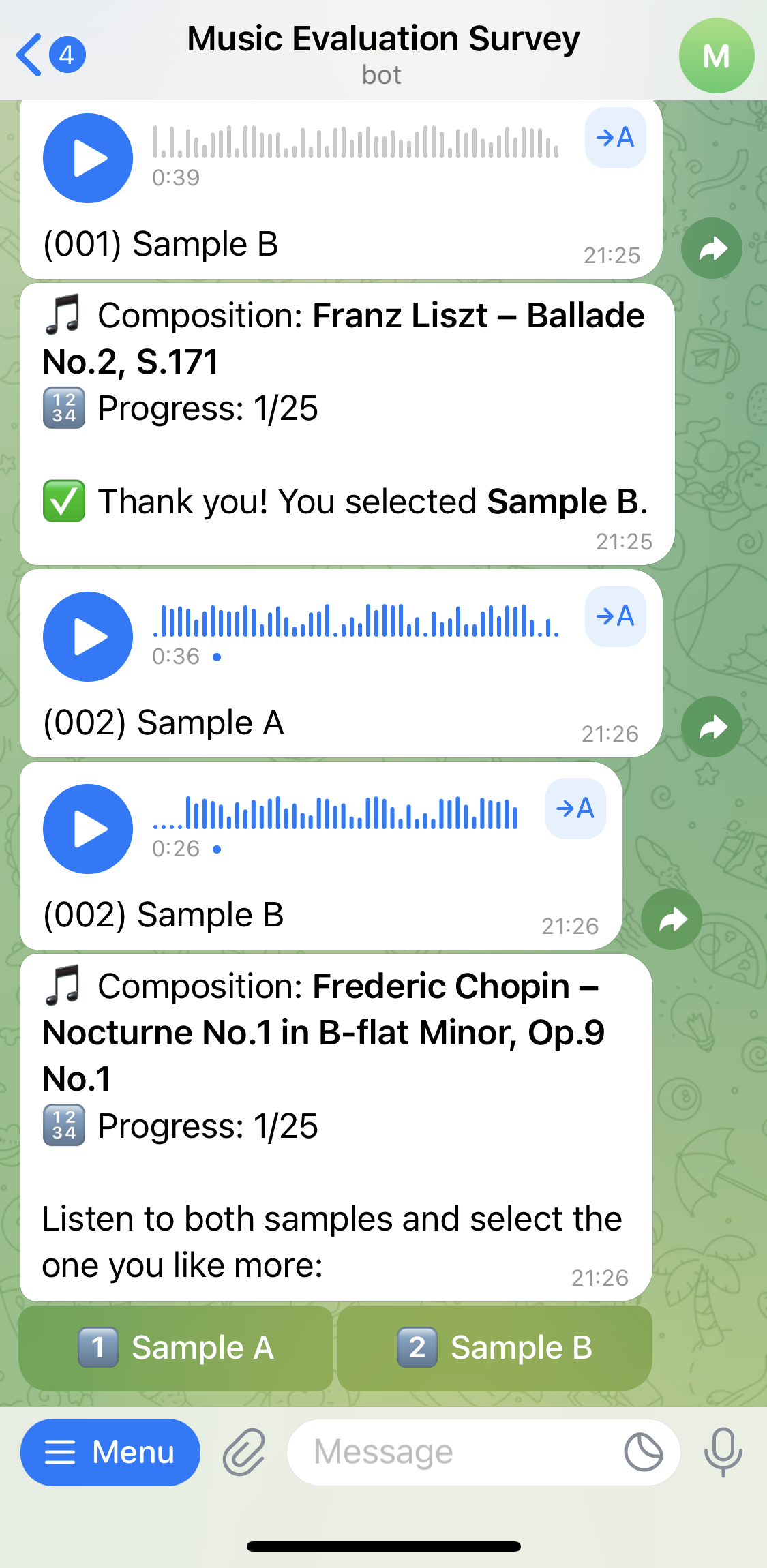}
        \caption{Feedback and the next issued pair of samples}
        \label{fig:sub4}
    \end{subfigure}
    \caption{Interface of the Telegram Bot developed for the side-by-side listening test.}
    \label{fig:bot}
\end{figure*}

\ifpdf\else
All of this shows that the results of listening tests depend on the selection and number of music pieces in the survey. A subset of the selected compositions can alter the overall rating. We advise readers to visit the \href{https://ilya16.github.io/SyMuPe}{demo page} and familiarize themselves with the samples produced by each model.

Finally, Figure~\ref{fig:bot} shows the interface of the Telegram bot developed for the evaluation survey. The process begins with a welcome message, followed by a series of paired ratings randomly issued by the bot. The sampling logic ensures that all compositions, models, and model pairs are uniformly rotated to obtain sufficient statistics. On average, each user saw each composition at least once and each pair of models at least twice.
\fi

\ifpdf
All of this shows that the results of listening tests depend on the selection and number of music pieces in the survey. A subset of the selected compositions can alter the overall rating. We advise readers to visit the \href{https://ilya16.github.io/SyMuPe}{demo page} and familiarize themselves with the samples produced by each model.

Finally, Figure~\ref{fig:bot} shows the interface of the Telegram bot developed for the evaluation survey. The process begins with a welcome message, followed by a series of paired ratings randomly issued by the bot. The sampling logic ensures that all compositions, models, and model pairs are uniformly rotated to obtain sufficient statistics. On average, each user saw each composition at least once and each pair of models at least twice.
\else\fi

\section{Ethical Statement}

All survey participants provided informed consent prior to participation. The study was conducted in accordance with applicable ethical guidelines, and all responses were anonymized before data analysis to ensure privacy and confidentiality. Some participants voluntarily provided individual responses and text feedback. The identities of the respondents were not disclosed.

This work uses audio recordings of piano performances found online to create symbolic piano performance MIDI datasets using an external audio-to-MIDI transcription model \cite{yan2024transkun}. To avoid copyright issues, we do not store or distribute any of the original audio files. This approach adheres to fair use principles, ensuring that the research remains focused on symbolic music representations for academic purposes.

% The derived MIDI data that forms the basis of our modeling and evaluation is abstract and does not reproduce copyrighted content. 

\end{document}